\documentclass[journal,draftclsnofoot,onecolumn,12pt,twoside]{IEEEtran}
\IEEEoverridecommandlockouts
\usepackage{amsmath}
\usepackage{amsfonts}
\usepackage{amssymb}
\usepackage{graphicx}
\usepackage{epstopdf}
\usepackage{caption}
\usepackage{subcaption}
\usepackage{cite}
\usepackage{algorithmicx}
\usepackage[ruled]{algorithm}
\usepackage{algpseudocode}
\usepackage{blindtext}
\usepackage{enumitem}
\usepackage{pdfpages}
\usepackage{bbm}
\usepackage{bibentry}
\usepackage{amsthm}

\usepackage{changes}
\definechangesauthor[color=blue,name={Mingyue Ji}]{MJ}

\ifodd 1

\else

\fi

\ifodd 1

\else

\fi
\renewcommand{\baselinestretch}{1.4}
\begin{document}
\title{Optimal Throughput-Outage Analysis of Cache-Aided Wireless Multi-Hop D2D Networks -- Derivations of Scaling Laws}

\author{Ming-Chun Lee,~\IEEEmembership{Student Member,~IEEE}, Mingyue Ji,~\IEEEmembership{Member,~IEEE}, and Andreas F. Molisch,~\IEEEmembership{Fellow,~IEEE}
\thanks{M.-C. Lee was with the Department of Electrical and Computer Engineering, University of Southern California, Los Angeles, CA 90089, USA. He is now with the Institute of Communications Engineering, National Chiao Tung University, Hsinchu 30010, Taiwan (email: mingchunlee@nctu.edu.tw).}
\thanks{A. F. Molisch is with the the Department of Electrical and Computer Engineering, University of Southern California, Los Angeles, CA 90089, USA (email: molisch@usc.edu).}
\thanks{M. Ji is with Department of Electrical and Computer Engineering, University of Utah, Salt Lake City, UT 84112, USA (email: mingyue.ji@utah.edu).}
\thanks{A condensed version of this paper has been submitted to IEEE Transactions on Communications \cite{lee2020optimal}. Part of this work will be presented in the 2020 IEEE Global Communications Conference \cite{lee2020throughput}.}
}


%Telephone: (800) 555--1212, Fax: (888) 555--1212}


\newcommand{\eqdef}{=\vcentcolon}

\maketitle
\vspace{-0.5cm}
\begin{abstract}
%\boldmath
Cache-aided wireless device-to-device (D2D) networks have demonstrated promising performance improvement for video distribution compared to conventional distribution methods. Understanding the fundamental scaling behavior of such networks is thus of paramount importance. However, existing  scaling laws for multi-hop networks have not been found to be optimal even for the case of Zipf popularity distributions (gaps between upper and lower bounds are not constants); furthermore, there are no scaling law results for such networks for the more practical case of a Mandelbrot-Zipf (MZipf) popularity distribution. We thus in this work investigate the throughput-outage performance for cache-aided wireless D2D networks adopting multi-hop communications, with the MZipf popularity distribution for file requests and users distributed according to Poisson point process. We propose an achievable content caching and delivery scheme and analyze its performance. By showing that the achievable performance is tight to the proposed outer bound, the optimal scaling law is obtained. Furthermore, since the Zipf distribution is a special case of the MZipf distribution, the optimal scaling law for the networks considering Zipf popularity distribution is also obtained, which closes the gap in the literature.
\end{abstract}
\vspace{-0.5cm}

\begin{IEEEkeywords}
Wireless caching network, device-to-device (D2D) communications, multi-hop D2D, throughput-outage analysis, scaling laws.
\end{IEEEkeywords}

% \fi
\IEEEpeerreviewmaketitle

\section{Introduction}

Due to the continuously increasing demand for video content delivery \cite{Cisco_2017}, finding new methods to provide highly efficient and low-cost video services is of paramount importance for 4G and 5G wireless systems \cite{Gol:femtocaching}. Conventional approaches \cite{Andrews:5G}, e.g., use of additional spectrum, massive antenna systems, and network densifications, all require obtaining more physical resources and/or deploying more infrastructure, either of which is costly. 

Over the last years, progress of semiconductor technology has made memory one of the cheapest hardware resources. Accordingly, caching at the wireless edge has emerged as a promising alternative approach for significantly improving the efficiency and quality of video distribution \cite{Gol:femtocaching}. The idea is to trade cheap memory resources for expensive bandwidth resources by caching video files close to the prospective users: the caches are filled during off-peak hours, and then provide the cached videos to the users requesting them in the peak hours. This principle, combined with the \emph {asynchronous content reuse} and \emph{concentrated popularity} that are a general characteristic of video requests \cite{Gol:femtocaching}, has rendered caching at the wireless edge a widely explored method for video distribution \cite{li2018survey}. It has been shown that caching methods can reduce the latency \cite{parvez2018survey}, increase the video delivery throughput and quality \cite{Choi:quality}, and reduce energy consumption \cite{lee2018cachingTWC}. Furthermore, information-theoretical studies have shown that caching methods can significantly improve the scaling laws of throughput \cite{Maddah-Ali:CCache,Ji:Th_Out_toff,Ji:Dcache,jeon2017wireless,lee2019throughput}.

There exist several different realizations of caching at the wireless edge. Among them, the most commonly discussed scenarios are caching at small cell base stations (BSs) with limited backhaul, e.g., femtocaching \cite{Gol:femtocaching,Shanmugam:fcache}, and caching at the user equipments (UE), e.g., cache-aided wireless device-to-device (D2D) communications \cite{Gol:femtocaching,Gol:Dcache1}. In either case, file delivery to the requesting users can be done either through simple transmission of complete files, or by a sophisticated joint encoding and multicasting of chunks from different files; the latter approach is known as coded-caching in the literature introduced in \cite{Maddah-Ali:CCache}. Recently, high performance D2D communication has been maturing \cite{Doppler:D2D}, and many papers have demonstrated that cache-aided D2D networks can improve various aspects of the video distribution without installing additional infrastructure \cite{ahmed2019video}. Moreover, the uncoded cache-aided D2D can provides the same throughput scaling law as the corresponding coded version \cite{Ji:Fund_D2D}, while maintaining the transmission simple. Thus, in this work, we focus on uncoded cache-aided wireless D2D networks.\footnote{Note that ``coding" here means the ``inter-file" coding. Error control coding within a file is still used.}

Numerous papers have reported results for uncoded cache-aided wireless D2D networks. They demonstrated that uncoded cache-aided D2D can successfully improve the network performance in terms of cache hit-rate \cite{Chen:Dcache}, throughput \cite{Ji:Dcache}, latency \cite{li2017delay}, energy efficiency (EE) \cite{EE:Chen}, and their tradeoffs \cite{lee2018cachingTWC,lee2019individual}, using either single-hop \cite{Ji:Th_Out_toff,Ji:Dcache} or multi-hop D2D communications \cite{wang2017multi,jeon2017wireless,ji2017multihop}. For example, caching policy designs were proposed to improve throughput and EE in \cite{lee2018cachingTWC}. Caching schemes that can provide low latency were proposed and analyzed in \cite{li2017delay} and \cite{amer2018inter}. Ref. \cite{Chen:Dcache} investigated the difference between cache hit-rate and throughput optimizations. In \cite{EE:Chen}, traffic offloading gain and energy cost were investigated when taking the battery life into consideration. In \cite{Ji:Dcache}, appealing throughput performance and robustness for cache-aided D2D networks were demonstrated with realistic propagation channels. Since results of cache-aided D2D networks have been published in hundreds of papers, the literature review in this paragraph cites only a sample of papers and topics. Comprehensive surveys on this area can be found in \cite{li2018survey,ahmed2019video,mehrabi2019device,prerna2020device}.

While most papers are devoted to improving the cache-aided D2D networks in practical setting, there exist papers that focus on understanding the fundamental properties and limits of cache-aided D2D networks \cite{gitzenis2012asymptotic,Ji:Th_Out_toff,Ji:Dcache,ji2017multihop,jeon2017wireless,lee2019performance,lee2019throughput,qiu2019popularity}. These papers use scaling law analysis to characterize how the network/user performance (e.g., throughput, outage, delay) scales as the number of users $n$ tends to infinity. Their results thus provide us with the performance trend as well as means of comparison between fundamentally different communication frameworks. This paper provides a contribution to this range of investigation. 

\subsection{Related Literature}

The throughput scaling law analysis\footnote{Scaling law order notations: given two functions $f$ and $g$, we say that: (1) $f(n)=\mathcal{O}(g(n))$ if there exists a constant $c$ and integer $N$ such that $f(n)\leq cg(n)$ for $n>N$. (2) $f(n)=o(g(n))$ if $\displaystyle{\lim_{n\to\infty}}\frac{f(n)}{g(n)}=0$. (3) $f(n)=\Omega(g(n))$ if $g(n)=\mathcal{O}(f(n))$. (4) $f(n)=\omega(g(n))$ if $g(n)=o(f(n))$. (5) $f(n)=\Theta(g(n))$ if $f(n)=\mathcal{O}(g(n))$ and $g(n)=\mathcal{O}(f(n))$.} for wireless D2D (or ad-hoc) networks has been subject to many investigations since the seminal work of Gupta and Kumar \cite{gupta2000capacity}. In \cite{gupta2000capacity}, the transport capacity was investigated under both a protocol model and a physical model, with multi-hop used for communications; $N$ users are either placed arbitrarily or randomly. Both a lower (achievable) bound on the throughput per user, of the order $\Theta\left(\frac{1}{\sqrt{N\log N}}\right)$ and an upper bound (under some conditions) of $\Theta\left(\frac{1}{\sqrt{N}}\right)$ were derived. In \cite{agarwal2004capacity}, a similar analysis was conducted with a generalized physical model and the upper bound $\Theta\left(\frac{1}{\sqrt{N}}\right)$ was validated under general conditions. The $\Theta(\sqrt{\log N})$ gap between the achievable throughput and the upper bound was closed in \cite{Franceschetti2007Closing}, however, with a slightly different model where the user distribution is described by a Poisson point process (PPP). A number of other schemes and channel models were investigated in other papers as well. For example, analysis involving fading effect was provided in \cite{xue2006scaling}. Also, the multicasting capacity was studied in \cite{shakkottai2010multicast} and \cite{niesen2010balanced}.

In parallel to the investigations of the throughput scaling law, scaling laws for the throughput-delay tradeoff in wireless ad-hoc networks were provided in \cite{gamal2004throughput,el2006optimal_I,el2006optimal_II}. In \cite{gamal2004throughput}, it was shown that the optimal throughput-delay tradeoff under random node distribution with no mobility is $D(N)=\Theta(T(N))$, where $D(N)$ and $T(N)$ are delay and throughput, respectively. Then, this result was generalized to networks with mobility in \cite{el2006optimal_I,el2006optimal_II}. Since all the previous investigations were based on multi-hop communications, a natural question is whether one can go beyond the scaling law bounds of multi-hop D2D communications by using more sophisticated physical layer processing. This was indeed shown to be the case in \cite{ozgur2007hierarchical}, which introduced a hierarchical cooperation scheme, where the cooperation between users is used to form a distributed multiple-input multiple-output (MIMO) system among the users and gain benefits. The resulting scaling of the throughput per user is {$\Theta\left(N^{-\varepsilon}\right)$ for arbitrarily small $\varepsilon$}, at the price of very complicated cooperation among user nodes.

Cache-aided D2D/ad-hoc networks have been substantially studied by the Computer Science community mostly with multi-hop communications, e.g., \cite{hara2001effective,ghandeharizadeh2004placement,yin2005supporting,hara2006data,cao2004cooperative,ghandeharizadeh2008cooperative}.  However, the fundamental scaling laws and optimality considerations did not draw much attention, except that \cite{cohen2002replication} proposed a caching policy (square-root replication policy) that provides the optimum design in terms of the {\em expected number of nodes} to visit until finding the desired content. Only recently did the fundamental properties of cache-aided D2D/ad-hoc networks start to draw more attention, and several papers have characterized the scaling laws of uncoded cache-aided D2D/ad-hoc networks. In \cite{golrezaei2014scaling}, the scaling law of the maximum expected throughput was characterized for single-hop cache-aided D2D networks considering a Zipf popularity distribution and a protocol model for transmission between nodes; however, it did not characterize the outage probability. {In fact, this maximum expected throughput can only be achieved when the outage probability goes to 1 as $N \rightarrow \infty$.}
To resolve this limitation, \cite{Ji:Th_Out_toff} investigated the scaling behavior of the throughput-outage performance for single-hop cache-aided D2D networks. It showed that the throughput per node can scale with {$\Theta\left(\frac{S}{M}\right)$} with negligibly small outage {probability} when a heavy-tailed Zipf popularity distribution is considered, where $M$ is the file library size and $S$ is the per-user memory size. This result was later generalized in \cite{lee2019throughput} by adopting the more practical and general modeling for the popularity distribution, namely the Mandelbrot-Zipf (MZipf) distribution. In \cite{gitzenis2012asymptotic}, the scaling law of the average throughput per node for the cache-aided D2D network with multi-hop communications was characterized with the assumption of user locations on a grid, while the tradeoff between throughput and outage was not explicitly investigated. Ref. \cite{jeon2017wireless} investigated the scaling law of the throughput-outage performance for cache-aided D2D networks with multi-hop communications, and provided an achievable throughput scaling law under the condition that the outage is vanishing. In \cite{qiu2019popularity}, an upper bound for the throughput scaling law was proposed, which complemented results in \cite{gitzenis2012asymptotic} and \cite{jeon2017wireless}. Notably, a major difference between the results in \cite{jeon2017wireless} and results in \cite{gitzenis2012asymptotic} and \cite{qiu2019popularity} was that \cite{jeon2017wireless} characterized the outage probability more explicitly.

There exist papers investigating scaling laws using more complicated delivery approaches. The scaling laws of coded cache-aided D2D/ad-hoc networks have been studied under different contexts, e.g., \cite{Ji:Fund_D2D,awan2015fundamental,naderializadeh2017fundamental,ibrahim2018device,yapar2019optimality}.
Besides, to improve cache-aided multi-hop D2D, schemes involving hierarchical cooperations were introduced in \cite{guo2017achievable,liu2018cache}, and their scaling laws were characterized. In contrast to the above papers which studied the scaling behavior of the throughput and outage performance, the scaling behavior of the throughput-delay tradeoff was studied in \cite{mahdian2017throughput}.

In this paper, we concentrate on the uncoded cache-aided D2D with simple multi-hop communications, i.e., the complicated hierarchical cooperation is not adopted. This is because the cache-aided multi-hop D2D has been shown to provide better scaling laws than the conventional unicasting, shared-link coded caching, and cache-aided single-hop D2D \cite{Ji:Th_Out_toff,lee2019throughput,Ji:Dcache,jeon2017wireless}. Also, the caching and delivery in D2D networks without coding can maintain the simplicity of the transmissions, while still providing the same scaling law as that with the corresponding coded version \cite{Ji:Fund_D2D}. As compared to schemes involving hierarchical cooperation, although those schemes could be better than those considering simple multi-hop schemes in some situations, their implementations are much more complicated. More importantly, the implementation of simple multi-hop D2D communications in wireless networks is more plausible thanks to the recent developments in ad-hoc networks \cite{al2014comprehensive} and D2D networks \cite{liu2014device,amodu2019primer}.

\subsection{Contributions}

In this work, we focus on the the scaling law analysis for the throughput-outage performance of uncoded cache-aided D2D with multi-hop communications. We aim to improve and complement results in previous papers \cite{jeon2017wireless,qiu2019popularity}. Specifically, when the outage is vanishing, \cite{jeon2017wireless} showed that the achievable throughput per user scales with $\Theta\left(\sqrt{\frac{S}{M\log(N)}}\right)$ for the heavy-tailed Zipf distribution, while the upper bound in \cite{jeon2017wireless} was $\Theta\left(\sqrt{\frac{S\log(N)}{M}}\right)$, where $S$ is the cache size of a user. Thus, there is a gap between the achievable performance and the upper bound. Ref. \cite{qiu2019popularity} provided a better throughput upper bound that scales as $\Theta\left(\sqrt{\frac{S}{M}}\right)$. However, similar to \cite{gitzenis2012asymptotic}, the tradeoff between throughput and outage performance was not characterized in \cite{qiu2019popularity}. Recently, based on a real-world dataset of mobile users, \cite{lee2019throughput} showed that, instead of the Zipf distribution adopted in \cite{jeon2017wireless} and \cite{qiu2019popularity}, a more general modeling for the popularity distribution is the MZipf distribution. Due to these observations, our paper thus aims to close the gap between the achievable throughput-outage performance and its outer bound as well as to provide scaling law analysis under the MZipf distribution assumption. Note that this paper is the first one to provide scaling law analysis for cache-aided wireless multi-hop D2D networks considering the MZipf popularity distribution. 

In this work, we use PPP to model the user distribution and use the MZipf distribution to model the popularity distribution of video requests \cite{lee2019throughput}. We assume the decentralized random caching policy \cite{Blaszczyszyn:fcache} and derive the tight achievable scaling laws of throughput-outage performance for the regimes that the outage performances are either negligibly small or converging to zero, corresponding to the practical requirement that the desirable outage of the network should be small. Our achievable scheme is obtained by first deriving the optimal caching policy, and then exploiting a hybrid clustering and multi-hop delivery scheme. We also provide the outer bound of the throughput-outage performance, again, for the regimes that the outage performance is either negligibly small or converging to zero. The outer bound is derived by analyzing the upper bound of the distances between the source-destination pairs.

We further show that the derived achievable {per-user throughput} scaling law and its outer bound are tight, i.e., the multiplicative gap between the lower and upper bounds can be upper bounded by a constant. Specifically, we show that when the outage performance is negligibly small, the throughput per user scales according to $\Theta\left(\sqrt{\frac{S}{M}}\right)$ for $\gamma<1$ and according to $\Theta\left(\sqrt{\frac{S}{q}}\right)$ for $\gamma>1$, where $\gamma$ is the Zipf factor and $q$ is the plateau factor of the MZipf distribution (see mathematical definition in Sec. II). Such result is intuitive as it indicates that, on the one hand, the performance is dominated by the file library size $M$ when we have a heavy-tailed popularity distribution. On the other hand, the performance is dominated by the plateau factor $q$, i.e., the total number of very popular files, when we have a light-tailed popularity distribution. We note that since the Zipf distribution is simply a special case of the MZipf distribution, as a by-product, our results close the gap of the Zipf distribution case in the literature, leading to the per user throughput scaling of $\Theta\left(\sqrt{\frac{S}{M}}\right)$ for $\gamma<1$ and almost $\Theta\left(\sqrt{S}\right)$ for $\gamma>1$. Moreover, since the multiplicative gap between the achievable scheme and the outer bound can be upper bounded by a constant, our achievable throughput-outage scaling law is optimum.

\subsection{Paper Organization}

The remainder of this paper is organized as follows. Sec. II discusses the network model, assumptions, and definitions for the throughput and outage of the network. Sec. III gives the proposed achievable scheme and the derived achievable scaling law. The outer bound is presented in Sec. IV. Since the results under the assumption of the standard Zipf distribution are important special cases, those results are presented independently in Sec. V. Conclusions are provided in Sec. VI. Proofs are relegated to Appendices.

\section{Network Setup}

We consider a random dense network where users are placed according to a PPP within a unit square-shaped area $[0,1]\times [0,1]$. We assume that the density of the PPP is $N$. As a result, the average number of users in the network is $N$ and the number of users $\mathsf{n}$ in the network is a random variable following the Poisson distribution.\footnote{Since our derivations are based on results in \cite{agarwal2004capacity} and \cite{Franceschetti2007Closing}, the results in this paper can be extended to the extended network in which users are placed according to a PPP with unit density in a square-shaped area of size $N$ (See \cite{agarwal2004capacity} and \cite{Franceschetti2007Closing}).} Accordingly, the probability that the network has $n$ users is:
\begin{equation}\label{eq:PoiDis}
\mathbb{P}_N(\mathsf{n}=n)=\frac{N^n}{n!}e^{-N}.
\end{equation} 
Note that according to the PPP, these $n$ users are uniformly distributed within the unit square area. Each device in the network can cache $S$ files. We consider a library consisting of $M$ files and assume that each file has equal size. We assume that users request the files from the library independently according to a request distribution modeled by the MZipf distribution \cite{Hefeeda:P2P,lee2019throughput}:
\begin{equation}
P_r(f;\gamma,q)=\frac{(f+q)^{-\gamma}}{\sum_{m=1}^M (m+q)^{-\gamma}}=\frac{(f+q)^{-\gamma}}{H(1,M,\gamma,q)},
\end{equation}
where $\gamma$ is the Zipf factor and $q$ is the plateau factor of the distribution; $H(a,b,\gamma,q):=\sum_{f=a}^b (f+q)^{-\gamma}$. We can see that the MZipf distribution degenerates to a Zipf distribution when $q=0$. To simplify the notation, we will in the remainder of this paper use $P_r(f)$ instead of $P_r(f;\gamma,q)$ as the short-handed expression. We consider the decentralized random caching policy for all users \cite{Blaszczyszyn:fcache}, in which users cache files independently according to the same caching policy. Denoting $P_c(f)$ as the probability that a user caches file $f$, the caching policy is fully described by $P_c(1),P_c(2),...,P_c(M)$, where $0\leq P_c(f)\leq 1,\forall f$; thus users cache files according to the caching policy $\lbrace P_c(f) \rbrace_{f=1}^M$. To satisfy the cache space constraint, we have $\sum_{f=1}^M P_c(f)=S$. In this paper, we assume that $S$ and $\gamma$ are some constants.

We consider the asymptotic analysis in this paper, in which we assume that $N\to\infty$ and $M\to\infty$. We will restrict to $M=o(N)$ and $q=\mathcal{O}(M)$ when $\gamma<1$; $M=o(N)$ and $q=o(M)$ when $\gamma>1$. The main reason for restricting to $M=o(N)$ when $\gamma<1$ is to render the users of the network the sufficient ability to cache the whole library. Similarly, the assumption that $q=o(M)$ and $M=o(N)$ when $\gamma>1$ renders the users of the network the sufficient ability to cache the most popular $q$ files (orderwise); otherwise the outage probability would go to $1$.

The plateau factor $q$ can either go to infinity or remain as a constant. When $q$ goes to infinity, it is sufficient to consider $q=\mathcal{O}(M)$. This is because the MZipf distribution would behave like a uniform distribution asymptotically as $q=\omega(M)$. As a result, we assume $q=\mathcal{O}(M)$ when $\gamma<1$. In addition, when $\gamma>1$, it is more interesting to consider the case that $q=o(M)$ because it gives a clear distinction between the heavy-tailed case ($\gamma<1$) and the light-tailed case ($\gamma>1$). The definition of a heavy-tailed popularity distribution can be found in Definition 3 of \cite{jeon2017wireless}. Note that if we have $q=\mathcal{O}(M)$ for the case that $\gamma>1$, due to impact of a large $q$, we would still have a heavy-tailed popularity distribution in this situation. As a result, we can expect that the scaling law in the case with $\gamma>1$ and $q=\Theta(M)$ to be similar to that with $\gamma<1$ and $q=\mathcal{O}(M)$, as the performance of the former case is restricted by $q$ \cite{lee2019throughput}. As a matter of practice, we see from the measurment results in \cite{lee2019throughput} that $q$ is much smaller than $M$ when $\gamma>1$, which supports the consideration of $q=o(M)$. When $q$ is a constant, i.e., $q=\Theta(1)$, the request distribution generally behaves like a Zipf distribution as $M\to\infty$. Thus, the results for $q=\Theta(1)$ can be representative for the analysis that uses the Zipf distribution for the request distribution.\footnote{We actually repeat all derivations for Zipf distribution, i.e., $q=0$, and find that this claim is true. The repeated derivations for Zipf distribution are omitted for brevity.} We will consider $q\to\infty$ in Secs. III and IV and consider $q=\Theta(1)$ in Sec. V.

We consider the physical model and define that the link rate between two users $i$ and $j$ follows the well-known physical model \cite{agarwal2004capacity,xue2006scaling}:
\begin{equation}\label{eq:physical_model}
R(i,j)=\left\{
\begin{aligned}
\log_2(1+\vartheta),\qquad&\log_2\left(1+\frac{P_il(i,j)}{N_0+\sum_{k\neq j}P_kl(k,j)}\right)\geq \vartheta\\
0,\qquad& \log_2\left(1+\frac{P_il(i,j)}{N_0+\sum_{k\neq j}P_kl(k,j)}\right)<\vartheta
\end{aligned}
\right.,
\end{equation}
where $\vartheta$ is some constant according to the delivery mechanism; $N_0$ is the noise power spectral density; $P_i$ is the power of user $i$; and {$l(i,j)=\frac{\chi}{d_{ij}^{\alpha}}$ is the power attenuation between users $i$ and $j$, where $d_{ij}$ is the distance between users $i$ and $j$, $\chi>0$ is some calibration factor,} and $\alpha>2$ is the pathloss coefficient.\footnote{It should be noted that when considering an extended network, we can extend our results in this paper to the bounded pathloss model, i.e., $l(i,j)=\min\left(1,\frac{\chi}{d_{ij}^{\alpha}}\right)$.} {It should be noticed that our scaling law analysis is based on the bounded rate model in (\ref{eq:physical_model}) instead of the Shannon capacity model. Therefore, as the rate is bounded, the fact that the pathloss model could be unbounded does not lead to unreasonable consequence for the scaling laws derived in this paper.} {Finally, we note that although this model will not be explicitly mentioned later, it is necessary for introducing the results of \cite{agarwal2004capacity} and \cite{xue2006scaling} (which are based on the same physical model), which help analyzing the achievable performance and outer bound.}

We consider multi-hop D2D delivery for the network. Users can only obtain their desired files through either multi-hop D2D delivery or self-caching. In other words, users can only obtain files from caches of the users in the network. Since $S$ is a constant, the probability that a user can find the desired file from its own cache goes to zero as $q$ and $M$ go to infinity. This prevents a possible gain from trivial self-caching; we thus concentrate on the analysis of D2D collaborative caching gain and assume without loss of generality that the throughput per user of using self-caching is identical to that using D2D-caching. Therefore, we do not distinguish between users retrieving the desired files from their own caches and from caches of other users. Moreover, similar to \cite{Ji:Th_Out_toff}, we assume that different users making the requests on the same file would request different segments of the file, which avoids the gain from the naive multicasting.

We define an outage as an occurrence that a user cannot obtain its desired file through either the multi-hop D2D delivery or self-caching. We define $\mathsf{n}$, $\mathsf{P}$, $\mathsf{F}$, and $\mathsf{G}$ as the random variables of number of users, location placement of users, file requests of users, and file placements of users. Then, suppose we are given a realization of number of users in the network with a realization of the placement of the user locations according to the binomial point process (BPP). In addition, we are given a realization of file requests and a realization of file placements of users according to the popularity distribution $P_r(\cdot)$ and caching policy $P_c(\cdot)$, respectively. We can define $T_u$ as the throughput of user $u\in\mathcal{U}$ under a feasible multi-hop file delivery scheme, where $\mathcal{U}$ is the user set when given the realization of $\mathsf{n}$. We then define the average throughput of user $u$ with a given number of users $n$ and location placement of users $r$ as $\overline{T}_u(n,r)=\mathbb{E}[T_u\mid \mathsf{n}=n,\mathsf{P}=r]$, where the expectation is taken over the file requests $\mathsf{F}$ of users, the file placement of users $\mathsf{G}$, and the file delivery scheme. Subsequently, we define 
\begin{equation}
T_{\text{user}}(n,r)=\displaystyle{\min_{u\in\mathcal{U}}\overline{T}_{u}(n,r)}.
\end{equation}
Finally, the expected minimum average throughput of a user in the network is defined as
\begin{equation}
\overline{T}_{\text{user}}=\mathbb{E}_{\mathsf{n},\mathsf{P}}[T_{\text{user}}(n,r)],
\end{equation}
where the expectation is taken over $\mathsf{n}$ and $\mathsf{P}$.

When the number of users in the network is $n$, we define
\begin{equation}
N_o(n)=\displaystyle{\sum_{u\in\mathcal{U}}}\mathbf{1}\lbrace\mathbb{E}[T_u\mid \mathsf{P},\mathsf{F},\mathsf{G}]=0\rbrace
\end{equation}
as the number of users that are in outage, where $\mathbf{1}\lbrace\mathbb{E}[T_u\mid \mathsf{P},\mathsf{F},\mathsf{G}]=0\rbrace$ is the indicator function such that the value is $1$ if $\mathbb{E}[T_u\mid \mathsf{P},\mathsf{F},\mathsf{G}]=0$; otherwise the value is $0$. Intuitively, $\mathbf{1}\lbrace\mathbb{E}[T_u\mid \mathsf{P},\mathsf{F},\mathsf{G}]=0\rbrace$ is equal to zero when the file delivery scheme cannot deliver the desired file to user $u$. We note that the expectation of $\mathbb{E}[T_u\mid \mathsf{P},\mathsf{F},\mathsf{G}]$ is taken over the file delivery scheme and $\mathbf{1}\lbrace\mathbb{E}[T_u\mid \mathsf{P},\mathsf{F},\mathsf{G}]=0\rbrace$ is a random variable with the distribution being the function of $\mathsf{P}$, $\mathsf{F}$, and $\mathsf{G}$. The outage probability in the case of $n$ users is then defined as
\begin{equation}
p_o(n)=\frac{1}{n}\mathbb{E}_{\mathsf{P},\mathsf{F},\mathsf{G}}[N_o(n)]=\frac{1}{n}\sum_{u\in\mathcal{U}}\mathbb{P}\left(\mathbb{E}[T_u\mid \mathsf{P},\mathsf{F},\mathsf{G}]=0\right).
\end{equation}
Consequently, the network outage probability is defined as
\begin{equation}\label{eq:MH_def_outage}
p_o=\mathbb{E}_{\mathsf{n}>0}[p_o(\mathsf{n}=n)]+\mathbb{P}_N(n=0).
\end{equation}
Note that since we consider $N\to\infty$, $\mathbb{P}_N(n=0)$ is actually negligible for the asymptotic analysis. In the following, we will aim to analyze the throughput-outage performance in terms of $\overline{T}_{\text{user}}$ and $p_o$. We will be especially interested in the regime that the outage probability $p_o$ is small, i.e., the regime that $p_o=\epsilon$, where $\epsilon$ is a negligibly small number or converges to zero. The main notations used in this paper are summarized in Table \ref{tb:1}.

Before introducing the technical results, it should be noted that our results are based on the asymptotic analysis, which assumes $N\to\infty$ and $M\to\infty$; our goal is to understand the fundamental properties and limits via such analysis. Thus, the optimality here is limited to asymptotic sense and the optimality with finite dimensional setup is beyond the scope of this paper. The main results of this paper are listed in Table \ref{tb:Results} and the intuitive explanations of the results are provided as ``Remark'' in the paper.

\begin{table}
{
\caption{Summary of the Main Notations}
\centering
\begin{tabular}{| c | c | c | c |}
\hline
Notations & Descriptions \\
\hline
$M$; $S$  & number of files in the library ; number of files that a user can cache \\
\hline
$P_r(f)$ ; $P_c(f)$ & request probability for file $f$ ; probability that a user cache file $f$ \\
\hline
$\gamma$ ; $q$ & Zipf factor; plateau factor of the MZipf distribution \\
\hline
$N$ ; $n$ ; $\mathsf{n}$ & density of the PPP ; realization of $\mathsf{n}$ ; number of users in the network as a random variable \\
\hline
$\mathbb{P}_N(n)$ ; $K$ & Poisson distribution with parameter $N$ ; reuse factor \\
\hline
$\mathsf{P}$ ; $\mathsf{F}$ ; $\mathsf{G}$ & placement of user locations ; file requests of users ; file placements of users \\
\hline
$T_u$ ; $\overline{T}_{\text{user}}$ & throughput of user $u$ ; average throughput of user, defined as $\overline{T}_{\text{user}}=\mathbb{E}[T_{\text{user}}(n,r)]$ \\
\hline
$\overline{T}_u(n,r)$ ; $T_{\text{user}}(n,r)$  & defined as $\overline{T}_u=\mathbb{E}[T_u\mid \mathsf{n}=n,\mathsf{P}=r]$ ; defined as $T_{\text{user}}(n,r)=\displaystyle{\min_{u\in\mathcal{U}}\overline{T}_{u}}$  \\
\hline
$p_o(n)$ ; $p_o$ & outage probability when the number of users in the network is $n$ ; network outage probability \\
\hline
$p_{\text{miss}}(n_{\text{s}})$ &  the probability that a user cannot find the desired file when $n_{\text{s}}$ number of users are searched \\
\hline
$g_c(M)$ & normalized size length of a cluster, namely, the cluster size \\
\hline
$P_c^*(\cdot)$; $C_1$ ; $C_2$ ; $z_f$ ; $\nu$ ; $m^*$ & optimum caching policy notations (see definitions in Theorem 1) \\
\hline
$\overline{L}(n,\mathsf{P})$ & average distance between transmitter and receiver in the network for a given $n$ \\
\hline
\end{tabular}
\label{tb:1}
}
\end{table}

\begin{table}
{
\caption{Scope and Applicability of the Main Results}
\centering
\begin{tabular}{| c | c | c | c |}
\hline
 & $\gamma$ & $q$  & Descriptions  \\
\hline
Lemma 1 & Arbitrary  & Arbitrary  & Outage probability expression \\
\hline
Theorem 1 & Arbitrary  & Arbitrary  & Proposed caching policy \\
\hline
Proposition 1 & $\gamma<1$ & $q\to\infty$, $q=\mathcal{O}(M)$ & Achievable outage probability \\
\hline
Proposition 3 & $\gamma>1$ & $q\to\infty$, $q=o(M)$ & Achievable outage probability \\
\hline
Theorem 2  and Corollary 1 & $\gamma<1$ & $q\to\infty$, $q=\mathcal{O}(M)$ & Achievable throughput-outage performance \\
\hline
Theorem 3 and Corollary 4 & $\gamma>1$ & $q\to\infty$, $q=o(M)$ & Achievable throughput-outage performance \\
\hline
Theorem 4 & $\gamma<1$ & $q\to\infty$, $q=\mathcal{O}(M)$ & Throughput-Outage performance outer bound \\
\hline
Theorem 5 & $\gamma>1$ & $q\to\infty$, $q=o(M)$ & Throughput-Outage performance outer bound \\
\hline
Theorems 6 -- 10 & $\gamma<1$, $\gamma>1$ & $q=\Theta(1)$ & Throughput-Outage performance for Zipf distributions \\
\hline
\end{tabular}
\label{tb:Results}
\vspace{-20pt}
}
\end{table}

\section{Achievable Throughput-outage Performance}

In this section, we derive the achievable throughput-outage performance of the network, in which we say $(T(P_o),P_o)$ is achievable if there exists a caching and multi-hop file delivery scheme such that $\overline{T}_{\text{user}}\geq T(P_o)$ and $p_o\leq P_o$. We will in the following first provide the achievable file delivery scheme, and then propose the achievable caching scheme. Accordingly, the achievable throughput-outage performance will be derived. In this section, we focus on the cases that $q\to\infty$, i.e., $q=\omega(1)$. The cases that $q=\Theta(1)$ will be discussed later in Sec. V.

\subsection{Achievable Caching and File Delivery Scheme}

\begin{figure}
\centering
\includegraphics[width=0.7\textwidth]{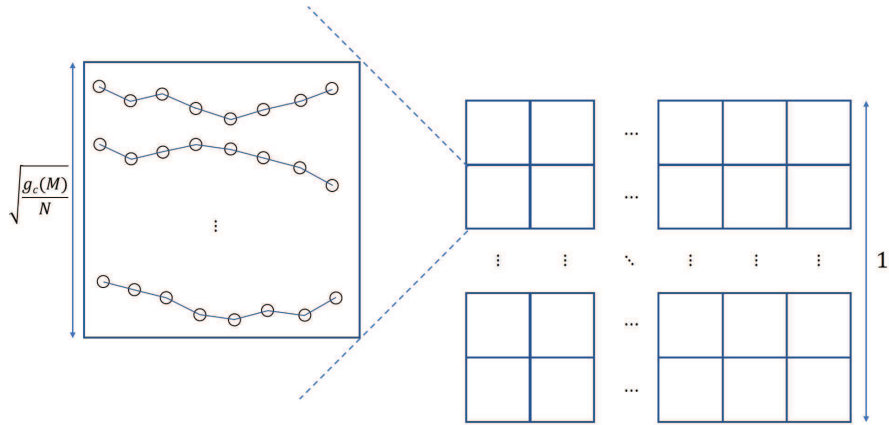}
\caption{Illustration for the achievable scheme in cache-aided multi-hop D2D networks.}
\label{fg:Fig_R3_Sys}
\vspace{-20pt}
\end{figure}

We consider the following achievable multi-hop file delivery scheme. We let $g_c(M)$ be a function of $M$ which goes to infinity as $M\to\infty$. Then, a clustering approach is used to split the cell into equally-sized square clusters, in which each cluster has the side length $\sqrt{\frac{g_c(M)}{N}}$, and $g_c(M)$ is thus denoted as the cluster size. Different clusters could be activated simultaneously. The inter-cluster interference is avoided by a Time Division Multiple Access (TDMA) scheme with reuse factor $K$ \cite{WireCom:Molisch}. Such a reuse scheme evenly applies $K$ colors to the clusters, and only the clusters with the same color can be activated on the same time-frequency resource for file delivery. We assume that a user in a cluster can only access files cached by users in the same cluster via either accessing its own cache or using (multi-hop) D2D communications following the multi-hop approach proposed in \cite{Franceschetti2007Closing}. The network model is illustrated in Fig. \ref{fg:Fig_R3_Sys}. Specifically, denoting $\mathcal{V}_f$ as the set of users in a cluster that cache file $f$, we consider the following transmission policy: for each user $u$ in the cluster, if the requested file $f$ can be found in the caches of users in $\mathcal{V}_f$, then a user $v_f$, randomly selected from $\mathcal{V}_f$, is set as the source for delivering the requested (real) file $f$ to user $u$; if the requested file cannot be found from the caches of any users in the cluster, user $u$ would be matched with a randomly selected user $v$ from users in the cluster, and then user $v$ is set as the source for delivering a {\em virtual} file to user $u$. Note that it does not matter what file is delivered in this case, as the user is actually in outage. 

After the establishment of the matching of the sources and destinations, to deliver (both real and virtual) files, the multi-hop approach proposed in \cite{Franceschetti2007Closing} directly applies. {Note that the delivery of virtual files cannot generate throughput for the network because users receiving virtual files are indeed in outage and the desired files are not actually received. However, we would still assume them to be included in the multi-hop D2D communications for the convenience of the mathematical analysis. Such scheme is suboptimal. Nevertheless, when the outage probability is either negligibly small or converging to zero, this scheme will be orderwise optimal because the performance degradation caused by delivering virtual files is negligible.} In addition, this delivery scheme can provide the per-user symmetric throughput for users that are not in outage because this is guaranteed by the adopted multi-hop approach in \cite{Franceschetti2007Closing}. Furthermore, since users cache files independently, the matching of the source-destination pairs here is equivalent to the uniformly random matching (see proof in Appendix O). Finally, we note that the assumption that a user may get a desired file from only its own cluster seems rather restrictive. However, the fact that this scheme can achieve (in the order sense) the outer bound shows that inclusion of inter-cluster communication cannot change the scaling law. 

By adopting the aforementioned scheme, due to the symmetry of the network and the thinning property of PPP, the throughput-outage performance for each cluster is the same as the throughput-outage performance for the whole network. We will thus in the following focus on the analysis of a cluster to derive $\overline{T}_{\text{user}}$ and $p_o$. In addition, since a user is in outage only if this user cannot find the desired file from any users in the same cluster, the outage probability $p_o$ is then equivalent to the probability that a user cannot find the desired file from users in the same cluster. Accordingly, when we denote the probability that a user can find the desired file in the cluster, i.e., the file hit-rate, as $P_h$, it is then clear that $P_h=1-p_o$.

To obtain the achievable caching scheme, we first provide Lemma 1 for the closed-form expression of $p_o$. Then, serving as the achievable caching scheme, the caching policy which minimizes $p_o$ is proposed in Theorem 1.

{\em Lemma 1:} Considering the proposed file delivery scheme, cluster size $g_c(M)$, and the caching distribution $P_c(\cdot)$, the outage probability of the proposed achievable scheme is
\begin{equation}
p_o=\sum_{f=1}^M P_r(f) e^{-g_c(M) P_c(f)}.
\end{equation}
\begin{proof}
See Appendix A.
\end{proof}

{\em Theorem 1:} Let $N\to\infty$, $M\to\infty$, $q\to\infty$, and $g_c(M)\to\infty$. Denote $m^*$ as the smallest index such that $P_c^*(m^*+1)=0$. Let $C_2=\frac{q\gamma}{S g_c(M)}$; $C_1$ is the solution of the equation: $C_1=1+C_2\log\left(1+\frac{C_1}{C_2}\right)$. The caching distribution $P_c^*(\cdot)$ that minimizes the outage probability $p_o$ is as follows:
\begin{equation}
P_c^*(f)=\left[\log\left(\frac{z_f}{\nu}\right) \right]^+,f=1,...,M,
\end{equation}
where $\nu=\exp\left(\frac{\sum_{f=1}^{m^*}\log z_f - S}{m^*}\right)$, $z_f=\left(P_r(f)\right)^{\frac{1}{g_c(M)}}$, $[x]^+=\max(x,0)$, and
\begin{equation}
m^*=\Theta\left(\min\left(\frac{C_1S g_c(M)}{\gamma},M \right)\right).
\end{equation}
\begin{proof}
See Appendix B.
\end{proof}

{\em Remark 1:} Similar to the results in \cite{lee2019throughput}, Theorem 1 indicates that the number of files with non-zero probability to be cached by users is at least on the same order as the plateau factor $q$ -- if $q=\mathcal{O}(g_c(M))$, then $m^*=\Theta(g_c(M))$; if $q=\omega(g_c(M))$, then $m^*=\Theta(q)$. This is intuitive when we look at the shape of the MZipf distribution: the most popular $q$ files (orderwise) have similar request probabilities, and we need to cache them to have the minimal outage probability.

{\em Remark 2:} Since Theorem 1 gives the optimal caching policy that minimizes the outage probability for a given cluster size and increasing cluster size can decrease the outage probability, these imply that the caching policy in Theorem 1 requires the smallest cluster size to achieve a given outage probability. Consequently, with an outage probability requirement for a {\em clustering network}, the caching policy in Theorem 1 can minimize the number of users in a cluster by minimizing the cluster size; this thus leads to the largest throughput per user.

Based on the achievable caching and file delivery scheme in this subsection, we subsequently characterize the achievable throughput-outage performance for both $\gamma<1$ and $\gamma>1$.

\subsection{Throughput-Outage Performance for $\gamma<1$}

In this subsection, we consider $\gamma<1$, $q=\omega(1)$, and $q=\mathcal{O}(M)$, and characterize the achievable throughput-outage performance. We will in the following first provide Proposition 1 to characterize the upper bound of the outage probability $p_o$. Then, Theorem 2 and Corollary 1 are provided to characterize the achievable throughput-outage performance. Finally, we use Proposition 2 and Corollary 2 to show that it is necessary to have $g_c(M)=\Theta(M)$ to obtain desirable outage probability when $\gamma<1$.

{\em Proposition 1:} Let $M\to\infty$, $N\to\infty$, and $q\to\infty$. Suppose $\gamma<1$ and let $D=\frac{q}{M}$. Consider $g_c(M)=\frac{\rho M}{C_1S}=o(N)$, where $\rho\geq \gamma$. When adopting the caching policy in Theorem 1, the outage probability $p_o$ is upper bounded as
\begin{equation}\label{eq:Corollay_2_2}
p_o\leq (1-\gamma)e^{-(\frac{\rho}{C_1}-\gamma)}\frac{D^{\gamma D}(1+D)^{-\gamma(1+D)}}{(1+D)^{1-\gamma}-D^{1-\gamma}}.
\end{equation}
\begin{proof}
See Appendix C.
\end{proof}

{\em Theorem 2:} Let $M\to\infty$, $N\to\infty$, and $q\to\infty$. Suppose $\gamma<1$ and let $D=\frac{q}{M}$. Consider $g_c(M)=\frac{\rho M}{C_1S}=o(N)$, where $\rho\geq \gamma$. When adopting the caching policy in Theorem 1, the following throughput-outage performance is achievable:
\begin{equation}
T(P_o)=\Omega\left(\frac{(1-P_o)}{K}\sqrt{\frac{C_1S}{\rho M}}\right), P_o=(1-\gamma)e^{-(\frac{\rho}{C_1}-\gamma)}\frac{D^{\gamma D}(1+D)^{-\gamma(1+D)}}{(1+D)^{1-\gamma}-D^{1-\gamma}}.
\end{equation}

\begin{proof}
See Appendix D.
\end{proof}

{\em Corollary 1:} Let $M\to\infty$, $N\to\infty$, and $q\to\infty$. Suppose $\gamma<1$ and $g_c(M)=\frac{\rho M}{C_1S}=o(N)$. When adopting the caching policy in Theorem 1 and considering $\rho=\Theta(1)\geq\gamma$, the following throughput-outage performance is achievable:
\begin{equation}
\begin{aligned}
&T(P_o)=\Omega\left(\sqrt{\frac{S}{\rho M}}\right),P_o=\epsilon_1(\rho),
\end{aligned}
\end{equation}
where $\epsilon_1(\rho)>0$ can be arbitrarily small. Furthermore, when considering $\rho\to\infty$, i.e., $\rho=\omega(1)$, we obtain the following achievable throughput-outage performance:
\begin{equation}
\begin{aligned}
&T(P_o)=\Omega\left(\sqrt{\frac{S}{\rho M}}\right),P_o=\Theta\left(e^{-\rho}\right)=o(1).
\end{aligned}
\end{equation}
\begin{proof}
Corollary 1 can be obtained directly from Theorem 2.
\end{proof}

{\em Remark 3:} From Theorem 2 and Corollary 1, we understand that when the outage probability is negligibly small, the achievable throughput is $\Omega\left(\sqrt{\frac{S}{M}}\right)$. In addition, Corollary 1 shows that when the outage probability converges to zero exponentially fast as $\rho \to \infty$, the achievable throughput is $\Omega\left(\sqrt{\frac{S}{\rho M}}\right)$. We will see later that this result meets the outer bound of the scaling law provided in Theorem 4 in Sec. IV.

{\em Proposition 2:} Let $M\to\infty$, $N\to\infty$, and $q\to\infty$. Suppose $\gamma\neq 1$ and $g_c(M)\to\infty$. Let $C_2=\frac{q\gamma}{S g_c(M)}$ and consider $g_c(M)<\frac{\gamma M}{C_1S}$. When adopting the caching policy in Theorem 1, the outage probability $p_o$ is:
\begin{equation}
\begin{aligned}
p_o= &1+(1-\gamma)e^{-\gamma\left(\frac{1}{C_1}-1\right)}\left(\frac{C_1S}{\gamma}\frac{g_c(M)}{M}\right)^{1-\gamma}\frac{\left(\frac{C_1}{C_1+C_2}\right)^{\gamma}\cdot\left(\frac{C_2}{C_1+C_2}\right)^{\gamma\frac{ C_2}{C_1}}}{\left(1+\frac{C_2S}{\gamma}\frac{g_c(M)}{M}\right)^{1-\gamma}-\left(\frac{C_2S}{\gamma}\frac{g_c(M)}{M}\right)^{1-\gamma}}\\
&\qquad - \left(\frac{C_1S}{\gamma}\frac{g_c(M)}{M}\right)^{1-\gamma}\frac{\left(1+\frac{C_2}{C_1}\right)^{1-\gamma}-\left(\frac{C_2}{C_1}\right)^{1-\gamma}}{\left(1+\frac{C_2S}{\gamma}\frac{g_c(M)}{M}\right)^{1-\gamma}-\left(\frac{C_2S}{\gamma}\frac{g_c(M)}{M}\right)^{1-\gamma}}.
\end{aligned}
\end{equation}

\begin{proof}
See Appendix F.
\end{proof}

{\em Corollary 2:} Let $M\to\infty$, $N\to\infty$, and $q\to\infty$. Suppose $\gamma<1$ and $g_c(M)\to\infty$. Consider $g_c(M)=o(M)$. When adopting the caching policy in Theorem 1, we obtain:
\begin{equation}
p_o= 1-o(1).
\end{equation}
\begin{proof}
This is proved by using Proposition 2 with $\gamma<1$ and $g_c(M)=o(M)$.
\end{proof}

{\em Remark 4:} Corollary 2 indicates that when $\gamma<1$, it is necessary to have $g_c(M)=\Theta(M)$ for guaranteeing the reasonable outage. Consequently, we are not interested in the cases that $g_c(M)=o(M)$.

\subsection{Throughput-Outage Performance for $\gamma>1$}

In this section, the achievable throughput-outage performance is characterized when $\gamma>1$, $q=\omega(1)$, and $q=o(M)$ are considered. In the following, we first use Proposition 3 and Corollary 3 to characterize the outage probability. Then, Theorem 3 and Corollary 4 are provided to characterize the achievable throughput-outage performance.

{\em Proposition 3:} Let $M\to\infty$, $N\to\infty$, and $q\to\infty$. Suppose $\gamma>1$ and $g_c(M)\to\infty$. Consider $g_c(M)=o(M)$ and $q=o(M)$. Let $C_2=\frac{q\gamma}{S g_c(M)}$. When adopting the caching policy in Theorem 1, the outage probability $p_o$ is:
\begin{equation}
\begin{aligned}
p_o=& 1+(\gamma-1)e^{-\gamma\left(\frac{1}{C_1}-1\right)}\cdot\left(\frac{C_1}{C_1+C_2}\right)^{\gamma}\cdot\left(\frac{C_2}{C_1+C_2}\right)^{\gamma\frac{ C_2}{C_1}}\cdot\left(\frac{C_2}{C_1}\right)^{\gamma-1}\\
&\qquad-\left(\left(\frac{C_1}{C_2}\right)^{\gamma-1}-\left(\frac{C_1}{C_1+C_2}\right)^{\gamma-1}\right)\cdot\left(\frac{C_2}{C_1}\right)^{\gamma-1}.
\end{aligned}
\end{equation}
\begin{proof}
See Appendix G.
\end{proof}

{\em Corollary 3:} Let $M\to\infty$, $N\to\infty$, and $q\to\infty$. Suppose $\gamma>1$ and $g_c(M)\to\infty$. Consider $g_c(M)=o(M)$, $q=o(M)$, and $g_c(M)=\frac{\alpha_1q}{S}$. When adopting the caching policy in Theorem 1 and considering $\alpha_1=\Theta(1)$, we can obtain $p_o=\epsilon_2(\alpha_1)$, where $\epsilon_2(\alpha_1)>0$ can be arbitrarily small. Furthermore, when $\alpha_1=\omega(1)$, i.e., $q=o(g_c(M))$, we obtain $p_o=\Theta\left(\frac{1}{(\alpha_1)^{\gamma-1}}\right)=o(1)$.
\begin{proof}
See Appendix H.
\end{proof}

\begin{figure}
\center
\includegraphics[width=0.7\textwidth]{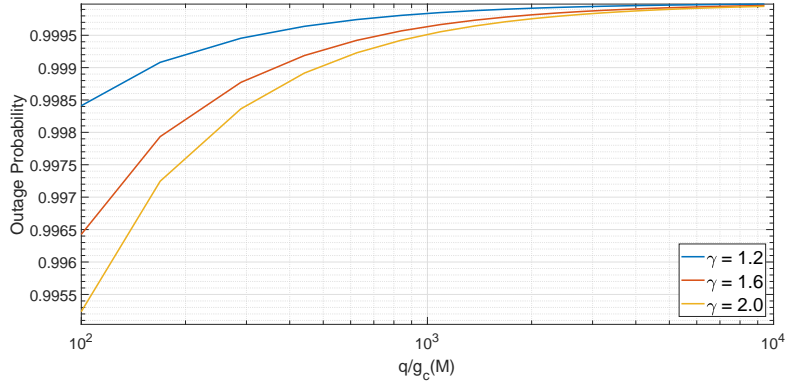}
\vspace{-10pt}
\caption{Outage probability with respect to $q/g_c(M)$ in Remark 5 considering different $\gamma$.}
\label{fg:Fig_D2D_Policy_Valid_Corollary_4}
\end{figure}

{\em Remark 5:} Let $M\to\infty$, $N\to\infty$, and $q\to\infty$. Suppose $\gamma>1$ and $g_c(M)\to\infty$. Consider $g_c(M)=o(M)$, $q=o(M)$, and $g_c(M)=\frac{\alpha_1q}{S}$. If we adopt the caching policy in Theorem 1 and gradually decrease $\alpha_1$, then $p_o(\alpha_1)$ would gradually increase. Note that this is simply a claim without rigorous proof being provided. However, this is very intuitive because increasing $\alpha_1$ is equivalent to decreasing $g_c(M)$. Besides, this can be observed by using Proposition 3 with computer simulations (See Fig. \ref{fg:Fig_D2D_Policy_Valid_Corollary_4}).

{\em Remark 6:} Remark 5 indicates that if $g_c(M)$ is not large enough for caching popular files in the plateau area of the MZipf distribution, the outage probability can be very large. On the contrary, Corollary 3 indicates that if $g_c(M)$ is large enough, the outage probability can go to zero as $\alpha_1\to\infty$. In practice, this implies that if $g_c(M)$ is large enough to cache the most popular $q$ files, we can have a reasonably good outage performance.

{\em Theorem 3:} Let $M\to\infty$, $N\to\infty$, and $q\to\infty$. Suppose $\gamma>1$ and $g_c(M)\to\infty$. Consider $g_c(M)=o(M)$, $q=o(M)$, and $g_c(M)=\frac{\alpha_1q}{S}$, where $\alpha_1=\Omega(1)$. When adopting the caching policy in Theorem 1, the following throughput-outage performance is achievable:
\begin{equation}
\begin{aligned}
T(P_o)&=\Omega\left(\frac{(1-P_o)}{K}\sqrt{\frac{S}{\alpha_1 q}}\right),\\
P_o&=1+(\gamma-1)e^{-\gamma\left(\frac{1}{C_1}-1\right)}\cdot\left(\frac{C_1}{C_1+C_2}\right)^{\gamma}\cdot\left(\frac{C_2}{C_1+C_2}\right)^{\gamma\frac{ C_2}{C_1}}\cdot\left(\frac{C_2}{C_1}\right)^{\gamma-1}\\
&\qquad-\left(\left(\frac{C_1}{C_2}\right)^{\gamma-1}-\left(\frac{C_1}{C_1+C_2}\right)^{\gamma-1}\right)\cdot\left(\frac{C_2}{C_1}\right)^{\gamma-1}.
\end{aligned}
\end{equation}

\begin{proof}
See Appendix I.
\end{proof}

{\em Corollary 4:} Let $M\to\infty$, $N\to\infty$, and $q\to\infty$. Suppose $\gamma>1$ and $g_c(M)\to\infty$. Consider $g_c(M)=o(M)$, $q=o(M)$, and $g_c(M)=\frac{\alpha_1 q}{S}$. When adopting the caching policy in Theorem 1 and considering $\alpha_1=\Theta(1)$ to be large enough, the following throughput-outage performance is achievable:
\begin{equation}
\begin{aligned}
&T(P_o)=\Omega\left(\sqrt{\frac{S}{\alpha_1 q}}\right),P_o=\epsilon_2(\alpha_1),
\end{aligned}
\end{equation}
where $\epsilon_2(\alpha_1)>0$ can be arbitrarily small. Furthermore, when considering $\alpha_1=\omega(1)\to\infty$, we obtain the following throughput-outage performance:
\begin{equation}
\begin{aligned}
&T(P_o)=\Omega\left(\sqrt{\frac{S}{\alpha_1 q}}\right),P_o=\Theta\left(\frac{1}{(\alpha_1)^{\gamma-1}}\right)=o(1).
\end{aligned}
\end{equation}
\begin{proof}
This is obtained by directly using Theorem 3 and Corollary 3.
\end{proof}

{\em Remark 7:} Theorem 3 and Corollary 4 characterize the achievable throughput-outage performance. Especially, Corollary 4 indicates that we can achieve the throughput $\Omega\left(\sqrt{\frac{S}{q}}\right)$ with a negligibly small outage probability. It also shows that when the outage probability converges to zero with the rate $(\alpha_1)^{\gamma-1}$, the achievable throughput is $\Omega\left(\sqrt{\frac{S}{\alpha_1 q}}\right)$. Besides, by comparing between Corollary 1 and Corollary 4, we understand that when the popularity distribution has a light tail, we can improve the scaling law: the performance is restricted by the order of $q$, instead of $M$. Finally, we will see that the achievable throughput-outage performance provided in Corollary 4 is optimum as it is tight to the outer bound provided in Theorem 5 in Sec. IV.

\subsection{Finite Dimensional Simulations}

In the subsection, we provide results of finite-dimensional simulations to compare the theoretical (solid lines) and simulated (dashed lines) curves of the achievable throughput-outage performance. The simulations are obtained by following the caching policy proposed in Theorem 1 and the delivery approach described in Sec. III.A. We assume that the routing is centrally controlled and the number of bits transmitted is sufficiently large such that there are bits to deliver by the network for the most of the time when obtaining the simulated results.

\begin{figure}
\mbox{
\hspace{-35pt}
\begin{subfigure}{0.55\textwidth}
\includegraphics[width=\textwidth]{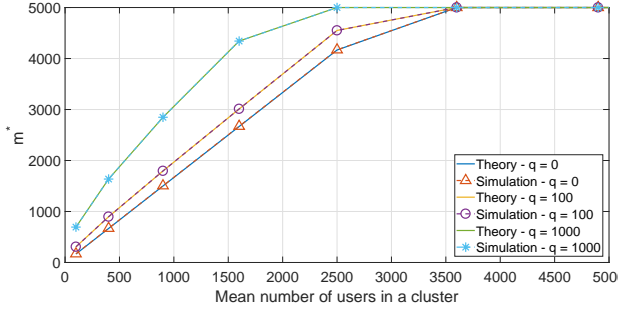}
\caption{$\gamma=0.6$, $S=1$, and $M=5000$.}
\end{subfigure}
\hspace{-30pt}
\begin{subfigure}{0.55\textwidth}
\includegraphics[width=\textwidth]{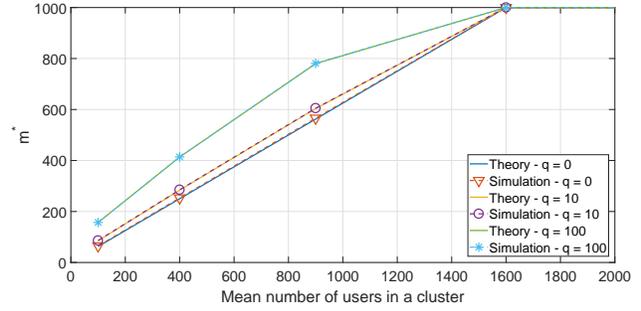}
\caption{$\gamma=1.6$, $S=1$, and $M=1000$.}
\end{subfigure}
}
\caption{Validations of Theorem 1 in terms of $m^*$.}
\label{fg:Fig_R3_1}
\end{figure}

\begin{figure}
\centering
\includegraphics[width=0.6\textwidth]{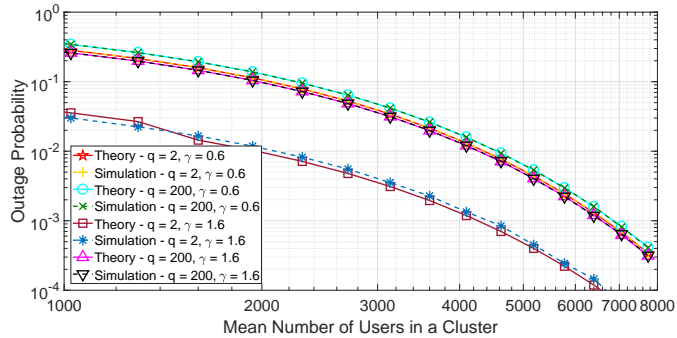}
\caption{Validations of the derived outage probability expressions in Propositions 1 and 3 with $S=1$ and $M=1000$.}
\label{fg:Fig_R3_2}
\end{figure}

\begin{figure}
\mbox{
\hspace{-20pt}
\begin{subfigure}{0.55\textwidth}
\includegraphics[width=\textwidth]{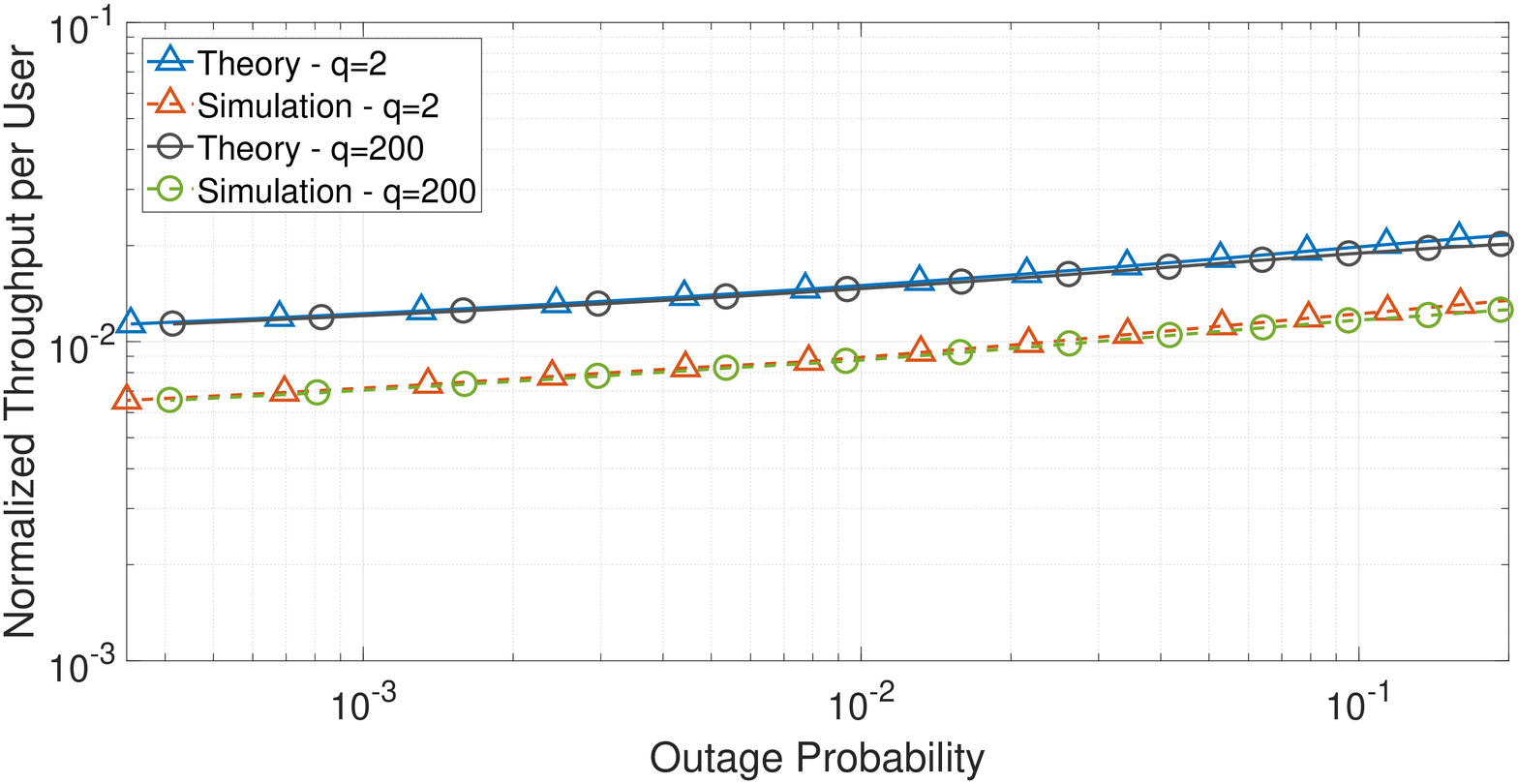}
\caption{$\gamma=0.6$.}
\end{subfigure}
\hspace{-30pt}
\begin{subfigure}{0.55\textwidth}
\includegraphics[width=\textwidth]{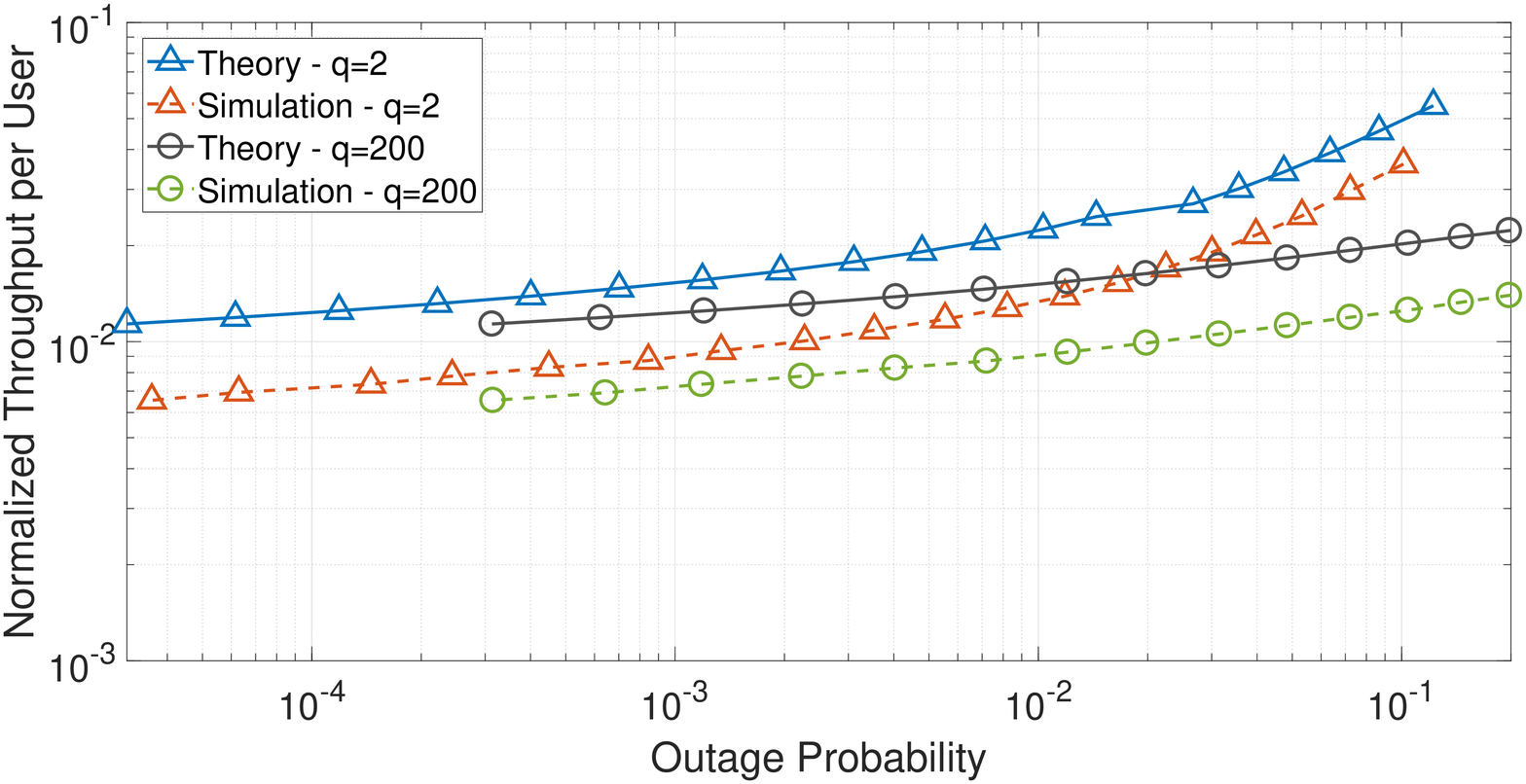}
\caption{$\gamma=1.6$.}
\end{subfigure}
}
\caption{Comparison between the normalized theoretical result (solid lines) and normalized simulated result (dashed lines) in networks adopting $S=1$ and $M=1000$.}
\label{fg:Fig_1}
\end{figure}

\begin{figure}
\centering
\includegraphics[width=0.6\textwidth]{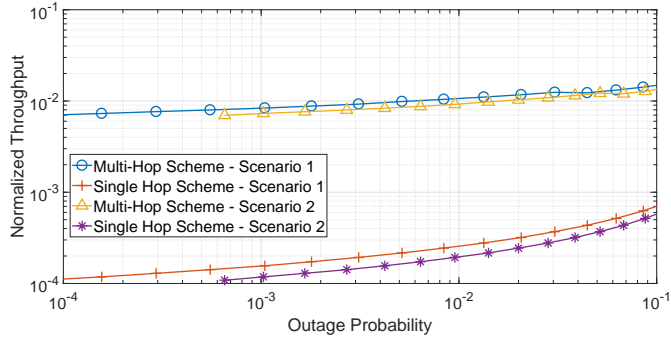}
\vspace{-10pt}
\caption{Performance comparisons between single-hop and multi-hop D2D networks. Parameters of Scenario 1 and Scenario 2 are in Table \ref{tb:R3_1}.}
\label{fg:Fig_R3_3}
\end{figure}

Specifically, we use Figs. \ref{fg:Fig_R3_1}-\ref{fg:Fig_1} to validate our proposed caching scheme and analysis and use Fig. \ref{fg:Fig_R3_3} to show performance. In Fig. \ref{fg:Fig_R3_1}, we validate the proposed caching policy in Theorem 1 in terms of $m^*$ by comparing the theoretical $m^*$ in Theorem 1 with the $m^*$ obtained by numerically solving the Karush-Kuhn-Tucker (KKT) conditions. The reason of using $m^*$ to validate Theorem 1 is that $m^*$ is the most critical parameter in Theorem 1. We observe that the theoretical results perfectly match the simulated results. We validate the proposed outage probability analysis in Fig. \ref{fg:Fig_R3_2}. Specifically, we compare the simulated results with the theoretical results in Propositions 1 and 3, i.e., the outage probability when $\gamma<1$ and $\gamma>1$, respectively. The results show that the simulated curves well match the theoretical curves.

In Figs. \ref{fg:Fig_1} and \ref{fg:Fig_R3_3}, we validate the proposed throughput-outage analysis and demonstrate the throughput-outage performance. The normalized throughput in the figures denotes the throughput that is normalized by the link capacity and effective reuse factor when using the D2D schemes. Fig. \ref{fg:Fig_1} shows that there is a constant factor gap (around $1.65$) between the theoretical and simulated curves. This might be caused by some constant factor between the theoretical throughput and simulated throughput results, as indicated in \cite{Franceschetti2007Closing,xue2006scaling}. However, such constant gap is minor when we consider the scaling law behavior which focuses on the order of the limiting case.\footnote{Note that this gap usually cannot be analytically quantized or characterized easily. The analytical characterization of such gap is beyond the scope of this paper.} In Fig. \ref{fg:Fig_R3_3}, we compare performance between the proposed multi-hop D2D network and a single-hop D2D network. The reason for this comparison is to numerically show that the cache-aided multi-hop D2D is much better than the cache-aided single-hop D2D, which matches the suggestion of the scaling law analysis. The single-hop D2D is based on the clustering approach proposed in \cite{lee2019throughput} with some modification which replaces the caching policy in \cite{lee2019throughput} with the caching policy proposed in Theorem 1 in this paper. Such modification is to match the fact that the user distribution in this paper is PPP. We adopt practical parameters of the MZipf distribution \cite{lee2019throughput} for simulations in Fig. \ref{fg:Fig_R3_3}. The parameters are shown in Table \ref{tb:R3_1}. Results show that the performance of the multi-hop D2D network is much better than the single-hop D2D network.

\begin{table}
\caption{Parameters of networks using the MZipf Model}
\centering
\begin{tabular}{| c | c | c | c | c || c | c | c | c | c |}
\hline
Scenario & $\gamma$ & $q$  & $M$ & $S$ & Scenario & $\gamma$ & $q$  & $M$ & $S$\\
\hline
1 & 1.16  & 22  & 7345 & 5 &
2 & 1.11  & 18  & 5405 & 5 \\
\hline
\end{tabular}
\label{tb:R3_1}
\end{table} 

\section{Outer Bound of the Throughput-outage Performance}

In this section, we derive the outer bound of the throughput-outage performance. In the following, we say a point $(T(P_o),P_o)$ is dominant (thus serving as an outer bound point) if, for any caching and delivery scheme, either $T(P_o)\geq \overline{T}_{\text{user}}$ or $P_o\leq p_o$ is satisfied. Note that although there are different dominant points, we will specifically characterize the dominant points where $P_o$ is either negligibly small or converging to zero. Besides, we again consider only the cases that $q\to\infty$ here. The cases that $q=\Theta(1)$ will be provided later in Sec. V.

{\em Theorem 4:} Let $M\to\infty$, $N\to\infty$, and $q\to\infty$. Suppose $\gamma<1$. When $\rho'=\Theta(1)$ is large enough, the throughput-outage performance of the network is dominated by:
\begin{equation}
\begin{aligned}
&T(P_o)= \Theta\left(\sqrt{\frac{S}{\rho' M}}\right),P_o= \epsilon_1'(\rho'),
\end{aligned}
\end{equation}
where $\epsilon_1'(\rho')>0$ can be arbitrarily small. Furthermore, when $\rho'=\omega(1)\to\infty$, the throughput-outage performance of the network is dominated by:
\begin{equation}
\begin{aligned}
&T(P_o)= \Theta\left(\sqrt{\frac{S}{\rho'M}}\right),P_o= \Theta(e^{-\rho'})=o(1).
\end{aligned}
\end{equation}
\begin{proof}
See Appendix J.
\end{proof}

{\em Theorem 5:} Let $M\to\infty$, $N\to\infty$, and $q\to\infty$. Suppose $\gamma>1$ and $q=o(M)$. When considering $\alpha_1'=\Theta(1)$, the throughput-outage performance of the network is dominated by:
\begin{equation}
T(P_o)=\Theta\left(\sqrt{\frac{S}{\alpha_1' q}}\right),P_o=\epsilon_2'(\alpha_1'),
\end{equation}
where $\epsilon_2'(\alpha_1')>0$ can be arbitrarily small. Furthermore, when considering $\alpha_1'=\mathcal{O}\left(q^{\frac{1}{\gamma-1}}\right)\to\infty$ but $\alpha_1'q=o(M)$, the throughput-outage performance of the network is dominated by:
\begin{equation}
\begin{aligned}
&T(P_o)=\Theta\left(\sqrt{\frac{S}{\alpha_1' q}}\right),P_o=\Theta\left(\frac{1}{(\alpha_1')^{\gamma-1}}\right)=o(1),
\end{aligned}
\end{equation}
\begin{proof}
See Appendix K.
\end{proof}

{\em Remark 8:} By comparing between Corollary 1 and Theorem 4, we observe that the achievable throughput-outage performance and the outer bound are tight to each other when $\gamma<1$. This indicates that the provided achievable scheme is orderwise optimal when the outage probability is either negligibly small or converging to zero.

{\em Remark 9:} By comparing between Corollary 4 and Theorem 5, we again see that there is no gap between the achievable throughput-outage performance and the outer bound when $\gamma>1$. This shows that the provided achievable scheme is orderwisely-optimal when the outage probability is either negligibly small or converging to zero.

\section{Main Results for $q=\Theta(1)$ Distributions}

In this section, we analyze the cases that $q=\Theta(1)$. It should be noted that since, asymptotically, a MZipf distribution with $q=\Theta(1)$ and $M\to\infty$ behaves equivalently to a Zipf distribution in terms of the throughput-outage performance, the results in this section can be representative for the results considering the Zipf distribution, i.e., $q=0$.\footnote{Note that the throughput-outage performance considering the standard Zipf distribution, i.e., $q=0$, can also be derived via repeating the approaches used in this paper with some algebraic differences on computations; these derivations will lead to the same results presented in the section.} Note that since proofs for Theorems 6, 7, and 9 can be done simply by setting $q=\Theta(1)$ and repeating the proofs for the corresponding theorems in Secs. III and IV, their proofs are omitted for simplicity.

The throughput-outage analysis results when considering $q=\Theta(1)$ are provided below. Theorem 6 provides the caching scheme used for deriving the achievable throughput-outage performance; Theorems 7 and 8 describe the achievable throughput-outage performance for $\gamma<1$ and $\gamma>1$, respectively; and Theorems 9 and 10 provide the throughput-outage outer bound for $\gamma<1$ and $\gamma>1$, respectively.

{\em Theorem 6:} Let $M\to\infty$, $q=\Theta(1)$, and $g_c(M)\to\infty$. Denote $m^*$ as the smallest index such that $P_c^*(m^*+1)=0$. The caching distribution $P_c^*(\cdot)$ that minimizes the outage is:
\begin{equation}
P_c^*(f)=\left[\log\left(\frac{z_f}{\nu}\right) \right]^+,f=1,...,M,
\end{equation}
where $\nu=\exp\left(\frac{\sum_{f=1}^{m^*}\log z_f - S}{m^*}\right)$, $z_f=\left(P_r(f)\right)^{\frac{1}{g_c(M)}}$, $[x]^+=\max(x,0)$, and
\begin{equation}
m^*=\Theta\left(\min\left(\frac{S g_c(M)}{\gamma},M \right)\right).
\end{equation}

{\em Theorem 7:} Let $M\to\infty$, $N\to\infty$, and $q=\Theta(1)$. Suppose $\gamma<1$ and consider $g_c(M)=\frac{\rho_{\text{zip}} M}{S}$, where $\rho_{\text{zip}}=\Omega(1)\geq \gamma$. When adopting the caching policy in Theorem 6 and considering $\rho_{\text{zip}}=\Theta(1)$, the following throughput-outage performance is achievable:
\begin{equation}
\begin{aligned}
&T(P_o)=\Omega\left(\frac{(1-P_o)}{K}\sqrt{\frac{S}{\rho_{\text{zip}} M}}\right), P_o=(1-\gamma)e^{-(\rho_{\text{zip}}-\gamma)}=\epsilon_{1,\text{zip}}(\rho_{\text{zip}}).\\
\end{aligned}
\end{equation}
Furthermore, when considering $\rho_{\text{zip}}\to\infty$, i.e., $\rho_{\text{zip}}=\omega(1)$, we obtain the following achievable throughput-outage performance:
\begin{equation}
\begin{aligned}
&T(P_o)=\Omega\left(\sqrt{\frac{S}{\rho_{\text{zip}} M}}\right),P_o=\Theta\left(e^{-\rho_{\text{zip}}}\right)=o(1).
\end{aligned}
\end{equation}

{\em Theorem 8:} Let $M\to\infty$ and $N\to\infty$, and $q=\Theta(1)$. Suppose $\gamma>1$. Consider $g_c(M)=o(M)$ and $g_c(M)=\frac{\alpha_{1,\text{zip}}' q}{S}$, where $\alpha_{1,\text{zip}}'=o(M)$ is any function that goes to infinity as $M\to\infty$. When adopting the caching policy in Theorem 1, the following throughput-outage performance is achievable:
\begin{equation}
\begin{aligned}
T(P_o)=\Omega\left(\sqrt{\frac{S}{\alpha_{1,\text{zip}}'}}\right),P_o=\Theta\left(\frac{1}{(\alpha_{1,\text{zip}}')^{\gamma-1}}\right)=o(1).
\end{aligned}
\end{equation}
\begin{proof}
See Appendix L.
\end{proof}

{\em Theorem 9:} Let $M\to\infty$, $N\to\infty$, and $q=\Theta(1)$. Suppose $\gamma<1$. When $\rho_{\text{zip}}'=\Theta(1)$ is large enough, the throughput-outage performance of the network is dominated by:
\begin{equation}
\begin{aligned}
&T(P_o)= \Theta\left(\sqrt{\frac{S}{\rho_{\text{zip}}'M}}\right),P_o= \epsilon_{1,\text{zip}}'(\rho_{\text{zip}}'),
\end{aligned}
\end{equation}
where $\epsilon_{1,\text{zip}}'(\rho_{\text{zip}}')>0$ can be arbitrarily small. Furthermore, when $\rho_{\text{zip}}'=\omega(1)\to\infty$ as $M\to\infty$, the throughput-outage performance of the network is dominated as:
\begin{equation}
\begin{aligned}
&T(P_o)= \Theta\left(\sqrt{\frac{S}{\rho_{\text{zip}}'M}}\right),P_o= \Theta(e^{-\rho_{\text{zip}}'}).
\end{aligned}
\end{equation}

{\em Theorem 10:} Let $M\to\infty$, $N\to\infty$, and $q=\Theta(1)$. Suppose $\gamma>1$. Suppose that $\alpha_{2,\text{zip}}'=o(M)\to\infty$ is any function that goes to infinity as $M\to\infty$. The throughput-outage performance of the network is dominated as:
\begin{equation}
\begin{aligned}
&T(P_o)=\Theta\left(\sqrt{\frac{S}{\alpha_{2,\text{zip}}'}}\right),P_o=\Theta\left(\frac{1}{(\alpha_{2,\text{zip}}')^{\gamma-1}}\right)=o(1).
\end{aligned}
\end{equation}
\begin{proof}
See Appendix M.
\end{proof}

{\em Remark 10:} By comparing Theorem 7 with Theorem 9, we observe that, the achievable throughput-outage performance and the outer bound are tight when $\gamma<1$. Likewise, by comparing Theorem 8 with Theorem 10, we see that the achievable throughput-outage performance and the outer bound are tight when $\gamma>1$.

With the results in this section, we can provide some comparisons with results in \cite{jeon2017wireless}, \cite{gitzenis2012asymptotic}, and \cite{qiu2019popularity}. We emphasize that although we tend to provide comparisons with results in those papers, different papers actually have slightly different network setups and assumptions; results in our paper are with more practical PPP model for the user distribution. Note that \cite{jeon2017wireless} and \cite{qiu2019popularity} adopted a simpler model in which the number of users in the network is a fixed value; \cite{gitzenis2012asymptotic} assumes that users are on a grid and communications can only happen between adjacent users on the grid. Besides, \cite{jeon2017wireless} provides detailed throughput-outage analysis for achievable performance with both $\gamma<1$ and $\gamma>1$ as well as the outer bound with $\gamma<1$, while the outer bound is not tight. Ref. \cite{gitzenis2012asymptotic} provides the achievable throughput scaling law without addressing the outer bound. In contrast, \cite{qiu2019popularity} provides only the outer bound without addressing the achievable performance. Notably, both \cite{gitzenis2012asymptotic} and \cite{qiu2019popularity} are with centralized caching policy and they do not consider the outage probability. We note that when compared with the centralized caching policy, the randomized policy, as being adopted in this paper, is easier to implement and more robust to user mobility \cite{Molisch:CachingEli}. The quantitative comparisons are remarked below.

{\em Remark 11:} By comparing Theorems 7 and 8 with results in \cite{jeon2017wireless}, we observe that our proposed achievable scheme improves the achievable throughput-outage performance. Specifically, the comparison results show that, when the outage converges to zero, the achievable throughput increases from $\Theta\left(\sqrt{\frac{S}{M\log(N)}}\right)$ to $\Theta\left(\sqrt{\frac{S}{\rho_{\text{zip}} M}}\right)$ when $\gamma<1$; and increases from $\Theta\left(\sqrt{\frac{S}{\log(N)}}\right)$ to $\Theta\left(\sqrt{\frac{S}{\alpha_{1,\text{zip}}'}}\right)$ for $\gamma>1$, where $\rho_{\text{zip}}$ and $\alpha_{1,\text{zip}}'$ can be any function such that $\rho_{\text{zip}},\alpha_{1,\text{zip}}'=\omega(1)$ as $M\to\infty$. In summary, our results improve the achievable throughput by almost $\mathcal{O}\left(\sqrt{\log(N)}\right)$ when the outage probability converges to zero.

{\em Remark 12:} By comparing Theorems 9 and 10 with the outer bound in \cite{jeon2017wireless}, we observe that our proposed outer bound improves the outer bound in \cite{jeon2017wireless} again by almost $\mathcal{O}\left(\sqrt{\log(N)}\right)$ for the case $\gamma<1$.\footnote{Theorem 10 actually provides an outer bound that characterizes the convergence rate with more details, as compared to the corresponding outer bound in \cite{jeon2017wireless}.}

{\em Remark 13:} When comparing our results with results in \cite{gitzenis2012asymptotic} and \cite{qiu2019popularity}, we see that our achievable throughput performance and outer bound are (orderwise) almost identical to the results provided in \cite{gitzenis2012asymptotic} and \cite{qiu2019popularity} when a negligibly small outage probability is allowed - they are with $\Theta\left(\sqrt{\frac{S}{M}}\right)$ for $\gamma<1$ and with virtually $\Theta\left(\sqrt{S}\right)$ for $\gamma>1$. The main difference between our results and results in \cite{gitzenis2012asymptotic} and \cite{qiu2019popularity} is that they have an additional regime for $1<\gamma<\frac{3}{2}$, where the throughput per node is worse than $\Theta\left(\sqrt{S}\right)$, whereas our results only have the regime $\gamma>1$. The possible reason might be that the allowance of a tiny outage (either negligibly small or converging to zero) in our results eliminate such regime. Note that since we consider the randomized caching policy, it is not possible to have an outage probability that is exactly zero, as opposed to the centralized caching strategies in \cite{gitzenis2012asymptotic} and \cite{qiu2019popularity}.

\section{Conclusions}

In this work, we conduct a scaling law analysis for the throughput-outage performance of the cache-aided D2D networks with multi-hop communications under the PPP and MZipf modeling for user distribution and popularity distribution, respectively. By demonstrating that there is no gap between the proposed achievable performance and outer bound, optimality is obtained in this work. Specifically, when $q=\omega(1)$, our results show that the optimal throughput per user scaling is $\Theta\left(\sqrt{\frac{S}{M}}\right)$ if $\gamma<1$ and $\Theta\left(\sqrt{\frac{S}{q}}\right)$ if $\gamma>1$. In addition, when $q=\Theta(1)$, our results show that the optimal throughput per user is $\Theta\left(\sqrt{\frac{S}{M}}\right)$ if $\gamma<1$ and almost $\Theta\left(\sqrt{S}\right)$ if $\gamma>1$. Note that all these results are either with negligibly small outage probability or with outage probability converging to zero, corresponding to the small outage requirement in practice. Since the analysis results for the case that $q=\Theta(1)$ can be representative for the results considering the standard Zipf distribution, our results close the gap between the achievable throughput-outage performance and outer bound that exists in the literature.

There are two possible future directions of this paper. First, as this paper focused on the regime that the outage probability is negligibly small or converging to zero, the throughput-outage tradeoff in other regimes have not been studies yet. The studies of other regimes are interesting because the tradeoffs might behave differently from the tradeoffs in this paper. Second, we assume in this paper that different users making the requests for the same file would request different segments of the file, which avoids multicasting opportunities. However, the multicasting opportunities might be possibly used for improving the scaling laws. Therefore, investigating whether and how involving multicasting opportunities can improve the current scaling laws could be a promising future direction.

\appendices

\section{Proof of Lemma 1}
According to (\ref{eq:PoiDis}), the probability of having $n$ users in a cluster is $\mathbb{P}_{g_c(M)}(n)$. Then, observe that $\mathbb{E}_{\mathsf{P},\mathsf{F},\mathsf{G}}\left[\frac{N_o(n)}{n}\right]$ is the probability that a user cannot find the desired file from the users in the cluster when there are $n$ users in a cluster. We thus obtain the outage probability:
\begin{equation}
\begin{aligned}
p_o&=\sum_{n=1}^{\infty} \mathbb{E}_{\mathsf{P},\mathsf{F},\mathsf{G}}\left[\frac{N_o(\mathsf{n})}{\mathsf{n}}\Big| \mathsf{n}=n\right]\mathbb{P}_{g_c(M)}(n)+ \mathbb{P}_{g_c(M)}(0)\\
&=\sum_{n=1}^{\infty}\sum_{f=1}^M  P_r(f)(1-P_c(f))^n\frac{(g_c(M))^n}{n!}e^{-g_c(M)} + e^{-g_c(M)}\\
&=\sum_{f=1}^M \sum_{n=1}^{\infty} P_r(f)(1-P_c(f))^n\frac{(g_c(M))^n}{n!}e^{-g_c(M)} + \sum_{f=1}^M P_r(f)e^{-g_c(M)}\\
&= \sum_{f=1}^M \sum_{n=0}^{\infty} P_r(f)(1-P_c(f))^n\frac{(g_c(M))^n}{n!}e^{-g_c(M)}\\
&=\sum_{f=1}^M \sum_{n=0}^{\infty} P_r(f)\frac{[g_c(M)(1-P_c(f))]^n}{n!}e^{-g_c(M)(1-P_c(f))}\cdot e^{-g_c(M)P_c(f)}\\
&=\sum_{f=1}^M P_r(f)e^{-g_c(M)P_c(f)} \underbrace{\sum_{n=0}^{\infty}\frac{[g_c(M)(1-P_c(f))]^n}{n!}e^{-g_c(M)(1-P_c(f))}}_{=1}=\sum_{f=1}^M P_r(f)e^{-g_c(M)P_c(f)}.
\end{aligned}
\end{equation}

\section{Proof of Theorem 1}

When considering a cluster with side length $\sqrt{\frac{g_c(M)}{N}}$, the number of users in the cluster is a Poisson random variable with mean equal to $g_c(M)$. Then according to Lemma 1, the optimization problem that minimizes the outage probability of the cluster is:
\begin{equation}
\begin{aligned}
\min &\quad\sum_{f=1}^M P_r(f) e^{-g_c(M) P_c(f)}\\
 s.t. &\quad\sum_{f=1}^M P_c(f) = S\\
 & 0\leq P_c(f)\leq 1,\forall f=1,2,...,M.
\end{aligned}
\end{equation}
Then since the optimization is convex, by using the Lagrange multiplier, we can obtain the optimal solution:
\begin{equation}
P_c^*(f)=\min\left( 1,\left[\frac{-1}{g_c(M)}\log\frac{\zeta}{g_c(M) P_r(f)} \right]^+ \right)=\min\left( 1,\left[\left(\log\frac{g_c(M) P_r(f)}{\zeta}\right)^{\frac{1}{g_c(M)}} \right]^+ \right),
\end{equation}
where $[a]^+=\max(a,0)$ and $\zeta$ is the Lagrange multiplier such that $\sum_{m=1}^M P_c^*(f)=S$.

To derive the final results of Theorem 1, in the following, we will first assume:
\begin{equation}\label{eq:Thm1_Assump}
P_c^*(f)=\min\left( 1,\left[\left(\log\frac{g_c(M) P_r(f)}{\zeta}\right)^{\frac{1}{g_c(M)}} \right]^+ \right)=\left[\left(\log\frac{g_c(M) P_r(f)}{\zeta}\right)^{\frac{1}{g_c(M)}} \right]^+.
\end{equation}
Then based on the resulting Theorem, we will show that the assumption in (\ref{eq:Thm1_Assump}) is indeed true for the caching policy derived in Theorem 1. Before starting the proof, we provide Lemma 2:

{Lemma 2:} Suppose $q\geq 0$ and $F>1$. We have the following inequalities:
\begin{equation}
\begin{aligned}
&\sum_{f=1}^F \log(f+q)\leq (F+q+1)\log(F+q+1)-F-(1+q)\log(1+q);\\
&\sum_{f=1}^F \log(f+q)\geq \log(1+q)+(F+q)\log(F+q)-F-(1+q)\log(1+q)+1.
\end{aligned}
\end{equation}
\begin{proof}
See Appendix E.
\end{proof}

We denote $\nu=\left(\frac{\zeta}{g_c(M)}\right)^{\frac{1}{g_c(M)}}$,  and $z_f=(P_r(f))^{\frac{1}{g_c(M)}}$. As a result, $P_c^*(f)=\left[\log\left(\frac{z_{f}}{\nu}\right)\right]^+$. We denote $m^*\leq M$ as the smallest index such that $P_c^*(m^*+1)=0$. Then since $P_c^*(f)$ is monotonically decreasing, we know that $\nu$ is a parameter such that $\log\left(\frac{z_{m^*+1}}{\nu}\right)>0$ and $\log\left(\frac{z_{m^*}}{\nu}\right)\leq0$. This leads to: $\frac{z_{m^*+1}}{\nu}>1$ and $\frac{z_{m^*}}{\nu}\leq 1$, i.e, $\nu< z_{m^*}$ and $\nu \geq z_{m^*+1}$. Observe that $\sum_{f=1}^{m^*} \log\left(\frac{z_f}{\nu}\right)=S$. It follows that
\begin{equation}
\sum_{f=1}^{m^*} \log\left(\frac{z_f}{z_{m^*}}\right)\leq S \quad ; \quad \sum_{f=1}^{m^*} \log\left(\frac{z_f}{z_{m^*+1}}\right)\geq S.
\end{equation}
As a result,
\begin{equation}\label{eq:Cond_nu}
\sum_{f=1}^{m^*} \log\left(\frac{P_r(f)}{P_r(m^*)}\right)^{\frac{1}{g_c(M)}}\leq S \quad ; \quad \sum_{f=1}^{m^*} \log\left(\frac{P_r(f)}{P_r(m^*+1)}\right)^{\frac{1}{g_c(M)}}\geq S.
\end{equation}
Recall that $P_r(f)=\frac{(f+q)^{-\gamma}}{H(1,M,\gamma,q)}$. It follows that
\begin{equation}\label{eq:Cond_nu_2}
\sum_{f=1}^{m^*} \log\left(\frac{P_r(f)}{P_r(m^*)}\right)^{\frac{1}{g_c(M)}}=\sum_{f=1}^{m^*}\log\left(\frac{f+q}{m^*+q}\right)^{\frac{-\gamma}{g_c(M)}}=\frac{-\gamma}{g_c(M)}\sum_{f=1}^{m^*}\log\left(\frac{f+q}{m^*+q}\right)
\end{equation}
By using Lemma 2, we know
\begin{equation}\label{eq:Lemma2_Thm_1}
\begin{aligned}
&\sum_{f=1}^{m^*} \log(f+q)\leq (m^*+q+1)\log(m^*+q+1)-m^*-(1+q)\log(1+q);\\
&\sum_{f=1}^{m^*} \log(f+q)\geq \log(1+q)+(m^*+q)\log(m^*+q)-m^*-(1+q)\log(1+q)+1.
\end{aligned}
\end{equation} 
As a result, we obtain:
\begin{equation}
\begin{aligned}
\frac{-\gamma}{g_c(M)}\sum_{f=1}^{m^*}\log\left(\frac{f+q}{m^*+q}\right)&\leq \frac{-\gamma}{g_c(M)}\left[\log(1+q)+(m^*+q)\log(m^*+q)-m^*-(1+q)\log(1+q)+1\right]\\
& \qquad + \frac{\gamma}{g_c(M)}m^*\log(m^*+q)\\
\frac{-\gamma}{g_c(M)}\sum_{f=1}^{m^*}\log\left(\frac{f+q}{m^*+q}\right)&\geq \frac{-\gamma}{g_c(M)}\left[(m^*+q+1)\log(m^*+q+1)-m^*-(1+q)\log(1+q)\right]\\
& \qquad + \frac{\gamma}{g_c(M)}m^*\log(m^*+q).
\end{aligned}
\end{equation}
This leads to
\begin{equation}\label{eq:UP_LP_1}
\begin{aligned}
\sum_{f=1}^{m^*} \log\left(\frac{P_r(f)}{P_r(m^*)}\right)^{\frac{1}{g_c(M)}}&\leq \frac{\gamma}{g_c(M)}\left[m^*-q\log(m^*+q)+q\log(1+q)-1\right]\\
\sum_{f=1}^{m^*} \log\left(\frac{P_r(f)}{P_r(m^*)}\right)^{\frac{1}{g_c(M)}}&\\
\geq \frac{\gamma}{g_c(M)}&\left[(m^*-(q+1)\log(m^*+q+1)-m^*\log\left(\frac{m^*+q+1}{m^*+q}\right)+(1+q)\log(1+q)\right].
\end{aligned}
\end{equation}
Similarly, we have 
\begin{equation}\label{eq:Cond_nu_2}
\sum_{f=1}^{m^*} \log\left(\frac{P_r(f)}{P_r(m^*+1)}\right)^{\frac{1}{g_c(M)}}=\sum_{f=1}^{m^*}\log\left(\frac{f+q}{m^*+q+1}\right)^{\frac{-\gamma}{g_c(M)}}=\frac{-\gamma}{g_c(M)}\sum_{f=1}^{m^*}\log\left(\frac{f+q}{m^*+q+1}\right)
\end{equation}
Hence, by using (\ref{eq:Lemma2_Thm_1}), we obtain:
\begin{equation}
\begin{aligned}
\frac{-\gamma}{g_c(M)}\sum_{f=1}^{m^*}\log\left(\frac{f+q}{m^*+q+1}\right)&\leq \frac{-\gamma}{g_c(M)}\left[\log(1+q)+(m^*+q)\log(m^*+q)-m^*-(1+q)\log(1+q)+1\right]\\
& \qquad + \frac{\gamma}{g_c(M)}m^*\log(m^*+q+1)\\
\frac{-\gamma}{g_c(M)}\sum_{f=1}^{m^*}\log\left(\frac{f+q}{m^*+q+1}\right)&\geq \frac{-\gamma}{g_c(M)}\left[(m^*+q+1)\log(m^*+q+1)-m^*-(1+q)\log(1+q)\right]\\
& \qquad + \frac{\gamma}{g_c(M)}m^*\log(m^*+q+1).
\end{aligned}
\end{equation}
By using (\ref{eq:Cond_nu_2}), this then leads to
\begin{equation}\label{eq:UP_LP_2}
\begin{aligned}
\sum_{f=1}^{m^*} \log\left(\frac{P_r(f)}{P_r(m^*+1)}\right)^{\frac{1}{g_c(M)}}&\leq \frac{\gamma}{g_c(M)}\left[m^*-q\log(m^*+q)-m^*\log\left(\frac{m^*+q}{m^*+q+1}\right)+q\log(1+q)-1\right]\\
\sum_{f=1}^{m^*} \log\left(\frac{P_r(f)}{P_r(m^*+1)}\right)^{\frac{1}{g_c(M)}}&\geq \frac{\gamma}{g_c(M)}\left[(m^*-(q+1)\log(m^*+q+1)+(1+q)\log(1+q)\right].
\end{aligned}
\end{equation}

We now let $a'=\frac{Sg_c(M)}{\gamma}$, $q=C_2a'$, and $m^*=C_1a'$, where $C_1$ and $C_2$ are some constant. We want to determine $m^*$ when $g_c(M)\to\infty$ (or equivalently $a'\to \infty$). From (\ref{eq:UP_LP_1}), we obtain :
\begin{equation}
\begin{aligned}\label{eq:app_opt_m_LB}
&\frac{1}{S}\sum_{f=1}^{m^*} \log\left(\frac{P_r(f)}{P_r(m^*)}\right)^{\frac{1}{g_c(M)}}\\
&\leq \frac{1}{a'}\left[C_1a'-C_2a'\log\left[(C_1+C_2)a'\right]+C_2a'\log(1+C_2a')-1\right]\\
&=C_1-C_2\log\left((C_1+C_2)a'\right)+C_2\log(1+C_2a')-\frac{1}{a'}\\
&=C_1-C_2\log\left[\frac{(C_1+C_2)a'}{C_2a'+1}\right]-\frac{1}{a'}=C_1-C_2\log\left[\frac{C_1+C_2}{C_2+\frac{1}{a'}}\right]-\frac{1}{a'}\\
&\approx C_1-C_2\log\left(1+\frac{C_1}{C_2}\right)
\end{aligned}
\end{equation}
Also from (\ref{eq:UP_LP_1}), we obtain:
\begin{equation}
\begin{aligned}\label{eq:app_opt_m_UB}
&\frac{1}{S}\sum_{f=1}^{m^*} \log\left(\frac{P_r(f)}{P_r(m^*)}\right)^{\frac{1}{g_c(M)}}\\
&\geq \frac{1}{a'}\left[C_1a'-(C_2a'+1)\log\left[(C_1+C_2)a'+1\right]-(C_1a')\log\left(\frac{(C_1+C_2)a'+1}{(C_1+C_2)a'}\right)+(1+C_2a')\log(1+C_2a')\right]\\
&=C_1-\left(C_2+\frac{1}{a'}\right)\log\left[(C_1+C_2)a'+1\right]-C_1\log\left(\frac{C_1+C_2+\frac{1}{a'}}{C_1+C_2}\right)+\left(C_2+\frac{1}{a'}\right)\log(1+C_2a')\\
&=C_1-\left(C_2+\frac{1}{a'}\right)\log\left[\frac{(C_1+C_2)a'+1}{C_2a'+1}\right]-C_1\log\left(\frac{C_1+C_2+\frac{1}{a'}}{C_1+C_2}\right)\\
&=C_1-\left(C_2+\frac{1}{a'}\right)\log\left[\frac{C_1+C_2+\frac{1}{a'}}{C_2+\frac{1}{a'}}\right]-C_1\log\left(\frac{C_1+C_2+\frac{1}{a'}}{C_1+C_2}\right)\\
&\approx C_1-C_2\log\left(1+\frac{C_1}{C_2}\right).
\end{aligned}
\end{equation}
As a result of (\ref{eq:app_opt_m_LB}) and (\ref{eq:app_opt_m_UB}), we obtain
\begin{equation}\label{app_opt_m_LUB}
C_1-C_2\log\left(1+\frac{C_1}{C_2}\right)\lessapprox\frac{1}{S}\sum_{f=1}^{m^*} \log\left(\frac{P_r(f)}{P_r(m^*)}\right)^{\frac{1}{g_c(M)}}\lessapprox C_1-C_2\log\left(1+\frac{C_1}{C_2}\right)
\end{equation}
Similarly, from (\ref{eq:UP_LP_2}), we can obtain:
\begin{equation}
\begin{aligned}\label{eq:app_opt_m_1_LB}
&\frac{1}{S}\sum_{f=1}^{m^*} \log\left(\frac{P_r(f)}{P_r(m^*+1)}\right)^{\frac{1}{g_c(M)}}\\
&\leq \frac{1}{a'}\left[C_1a'-C_2a'\log\left[(C_1+C_2)a'\right]-(C_1a')\log\left(\frac{(C_1+C_2)a'}{(C_1+C_2)a'+1}\right)+C_2a'\log(1+C_2a')-1\right]\\
&=C_1-C_2\log\left((C_1+C_2)a'\right)-C_1\log\left(\frac{(C_1+C_2)a'}{(C_1+C_2)a'+1}\right)+C_2\log(1+C_2a')-\frac{1}{a'}\\
&=C_1-C_2\log\left[\frac{(C_1+C_2)a'}{C_2a'+1}\right]-C_1\log\left(\frac{C_1+C_2}{C_1+C_2+\frac{1}{a'}}\right)-\frac{1}{a'}\\
&\approx C_1-C_2\log\left(1+\frac{C_1}{C_2}\right)
\end{aligned}
\end{equation}
Again from (\ref{eq:UP_LP_2}), we obtain:
\begin{equation}
\begin{aligned}\label{eq:app_opt_m_1_UB}
&\frac{1}{S}\sum_{f=1}^{m^*} \log\left(\frac{P_r(f)}{P_r(m^*+1)}\right)^{\frac{1}{g_c(M)}}\\
&\geq \frac{1}{a'}\left[C_1a'-(C_2a'+1)\log\left[(C_1+C_2)a'+1\right]+(1+C_2a')\log(1+C_2a')\right]\\
&=C_1-\left(C_2+\frac{1}{a'}\right)\log\left[(C_1+C_2)a'+1\right]+\left(C_2+\frac{1}{a'}\right)\log(1+C_2a')\\
&=C_1-\left(C_2+\frac{1}{a'}\right)\log\left[\frac{(C_1+C_2)a'}{C_2a'+1}\right]\\
&\approx C_1-C_2\log\left(1+\frac{C_1}{C_2}\right).
\end{aligned}
\end{equation}
As a result of (\ref{eq:app_opt_m_1_LB}) and (\ref{eq:app_opt_m_1_UB}), we obtain
\begin{equation}\label{app_opt_m_1_LUB}
C_1-C_2\log\left(1+\frac{C_1}{C_2}\right)\lessapprox\frac{1}{S}\sum_{f=1}^{m^*} \log\left(\frac{P_r(f)}{P_r(m^*+1)}\right)^{\frac{1}{g_c(M)}}\lessapprox C_1-C_2\log\left(1+\frac{C_1}{C_2}\right).
\end{equation}
Finally, by using (\ref{eq:Cond_nu}), (\ref{app_opt_m_LUB}), and (\ref{app_opt_m_1_LUB}), we obtain the following relationship: $C_1-C_2\log\left(1+\frac{C_1}{C_2}\right)=1$. Recall that $a'=\frac{Sg_c(M)}{\gamma}$, $C_2=\frac{a'}{q}$, and $m^*=C_1a'$. We conclude that
\begin{equation}
m^*=\Theta\left(\min\left(\frac{C_1Sg_c(M)}{\gamma},M \right)\right),
\end{equation}
where $C_1$ is the solution of $C_1=1+C_2\log\left(1+\frac{C_1}{C_2}\right)$.

Since above derivations are based on the assumption in (\ref{eq:Thm1_Assump}), we now show that this assumption is indeed true for the caching policy in Theorem 1. Observe that since $P_c^*(f)$ is a monotonically decreasing function of $f$, it is sufficient to show $P_c^*(1)\leq 1$. Then since $P_c^*(1)=\log\left(\frac{z_{1}}{\nu}\right)$, this is equivalent to show $\log z_1-\log\nu\leq 1$ as in the following. We first observe from above results that
\begin{equation}
\sum_{f=1}^{m^*} \log \left(\frac{z_f}{\nu}\right) = S<=>\sum_{f=1}^{m^*} \log z_f - m^*\log\nu = S
\end{equation}
It follows that 
\begin{equation}\label{eq:AppA_1}
-\log\nu = \frac{-1}{m^*}\sum_{f=1}^{m^*} \log z_f +\frac{S}{m^*}.
\end{equation}
Then, observe that 
\begin{equation}
\begin{aligned}\label{eq:AppA_1_1}
\sum_{f=1}^{m^*}\log z_f&=\sum_{f=1}^{m^*} \log (P_r(f))^{\frac{1}{g_c(M)}}= \frac{-\gamma}{g_c(M)}\sum_{f=1}^{m^*}\log (f+q) - \frac{1}{g_c(M)}\sum_{f=1}^{m^*}\log H(1,M,\gamma,q) \\
&= \frac{-\gamma}{g_c(M)}\sum_{f=1}^{m^*}\log (f+q) - \frac{m^*}{g_c(M)}\log H(1,M,\gamma,q).
\end{aligned}
\end{equation}
It follows from (\ref{eq:AppA_1}) and (\ref{eq:AppA_1_1}) that
\begin{equation}\label{eq:AppA_2}
-\log\nu = \frac{1}{m^*}\frac{\gamma}{g_c(M)}\sum_{f=1}^{m^*}\log (f+q) + \frac{1}{g_c(M)}\log H(1,M,\gamma,q) + \frac{S}{m^*}.
\end{equation}
By using Lemma 2, we obtain:
\begin{equation}\label{eq:AppA_3}
\sum_{f=1}^{m^*}\log (f+q)\leq \left(m^*+q+1\right)\log\left(m^*+q+1\right)-m^*-(1+q)\log(1+q).
\end{equation}
By using (\ref{eq:AppA_2}) and (\ref{eq:AppA_3}), we then obtain
\begin{equation}\label{eq:AppA_4}
\begin{aligned}
-\log\nu\leq &\frac{\gamma}{g_c(M)}\frac{1}{m^*}\left[\left(m^*+q+1\right)\log\left(m^*+q+1\right)-m^*-(1+q)\log(1+q)\right]\\
& + \frac{1}{g_c(M)}\log H(1,M,\gamma,q) + \frac{S}{m^*}.
\end{aligned}
\end{equation}
Recall that
\begin{equation}\label{eq:AppA_z1}
\log z_1=\log (P_r(1))^{\frac{1}{g_c(M)}}=\frac{1}{g_c(M)}\log \left((1+q)^{-\gamma}\right)-\frac{1}{g_c(M)}\log H(1,M,\gamma,q).
\end{equation}
By using (\ref{eq:AppA_4}) and (\ref{eq:AppA_z1}), we obtain:
\begin{equation}
\begin{aligned}\label{eq:Assumpt_Thm_1_0}
&\log z_1-\log\nu\\
&\leq \frac{\gamma}{g_c(M)}\frac{1}{m^*}\left[\left(m^*+q+1\right)\log\left(m^*+q+1\right)-m^*-(1+q)\log(1+q)\right]\\
& \qquad + \frac{1}{g_c(M)}\log H(1,M,\gamma,q) + \frac{S}{m^*}+\frac{1}{g_c(M)}\log \left((1+q)^{-\gamma}\right)-\frac{1}{g_c(M)}\log H(1,M,\gamma,q)\\
&= \frac{\gamma}{g_c(M)}\left[\left(1+\frac{q}{m^*}+\frac{1}{m^*}\right)\log\left(m^*+q+1\right)-1-\left(\frac{q}{m^*}+\frac{1}{m^*}\right)\log(1+q)\right]+ \frac{S}{m^*}-\frac{\gamma}{g_c(M)}\log(1+q).
\end{aligned}
\end{equation}

To show $\log z_1-\log\nu\leq 1$, we discuss two cases: (i) $\frac{C_1Sg_c(M)}{\gamma}\leq M$ and (ii) $\frac{C_1Sg_c(M)}{\gamma}> M$. We first consider $\frac{C_1Sg_c(M)}{\gamma}\leq M$. Then noticing that in this case $m^*=\frac{C_1Sg_c(M)}{\gamma}$. In addition, we have $q=C_2a'=\frac{C_2Sg_c(M)}{\gamma}$ and $g_c(M)\to\infty$. It follows from (\ref{eq:Assumpt_Thm_1_0}) that
\begin{equation}\label{eq:Assumpt_Thm_1_a}
\begin{aligned}
&\log z_1-\log\nu\leq \frac{\gamma}{g_c(M)}\cdot\\
&\left[\left(1+\frac{C_2}{C_1}+\frac{\gamma}{C_1Sg_c(M)}\right)\log\left(\frac{(C_1+C_2)Sg_c(M)}{\gamma}+1\right)-1-\left(\frac{C_2}{C_1}+\frac{\gamma}{C_1Sg_c(M)}\right)\log\left(1+\frac{C_2Sg_c(M)}{\gamma}\right)\right]\\
& \qquad+\frac{\gamma}{C_1g_c(M)}-\frac{\gamma}{g_c(M)}\log\left(1+\frac{C_2Sg_c(M)}{\gamma}\right)\\
&=\frac{\gamma}{g_c(M)}\left[\left(1+\frac{\gamma}{C_1Sg_c(M)}\right)\log\left(\frac{\frac{(C_1+C_2)Sg_c(M)}{\gamma}+1}{\frac{C_2Sg_c(M)}{\gamma}+1}\right)-1\right]\\
&\qquad+\frac{\gamma}{g_c(M)}\frac{C_2}{C_1}\log\left(\frac{\frac{(C_1+C_2)SNg_c(M)}{\gamma}+1}{\frac{C_2Sg_c(M)}{\gamma}+1}\right)+\frac{\gamma}{C_1g_c(M)}\\
&=\frac{\gamma}{g_c(M)}\left[\left(1+\frac{\gamma}{C_1Sg_c(M)}\right)\log\left(1+\frac{C_1}{C_2}\right)-1\right]+\frac{\gamma}{g_c(M)}\frac{C_2}{C_1}\log\left(1+\frac{C_1}{C_2}\right)+\frac{\gamma}{C_1g_c(M)}+o(1).
\end{aligned}
\end{equation}
Recall that $C_1=1+C_2\log\left(1+\frac{C_1}{C_2}\right)$ according to Theorem 1. We thus have:
\begin{equation}\label{eq:Assumpt_Thm_1_b}
\log\left(1+\frac{C_1}{C_2}\right)=\frac{C_1-1}{C_2}.
\end{equation}
By combining (\ref{eq:Assumpt_Thm_1_a}) and (\ref{eq:Assumpt_Thm_1_b}), we finally obtain:
\begin{equation}\label{eq:Assumpt_Thm_1_I}
\begin{aligned}
\log z_1-\log\nu&\leq\frac{\gamma}{g_c(M)}\left[\left(1+\frac{\gamma}{C_1Sg_c(M)}\right)\frac{C_1-1}{C_2}-1\right]+\frac{\gamma}{g_c(M)}\frac{C_2}{C_1}\frac{C_1-1}{C_2}+\frac{\gamma}{C_1g_c(M)}+o(1)\\
&=\frac{\gamma}{g_c(M)}\left[\frac{C_1}{C_2}-\frac{1}{C_2}\right]+\frac{\gamma}{g_c(M)}\frac{\gamma}{C_1Sg_c(M)}\frac{C_1-1}{C_2}+o(1)\stackrel{(a)}{=}o(1),
\end{aligned}
\end{equation}
where $(a)$ is because $N\to\infty$ and $\frac{C_1}{C_2}=\Theta(1)$ according to the equation $C_1=1+C_2\log\left(1+\frac{C_1}{C_2}\right)$. We now consider $\frac{C_1Sg_c(M)}{\gamma}> M$. In this case, we have $m^*=M$. Then from (\ref{eq:Assumpt_Thm_1_0}), we obtain:
\begin{equation}
\begin{aligned}\label{eq:Assumpt_Thm_1_II}
&\log z_1-\log\nu\\
&\leq \frac{\gamma}{g_c(M)}\left[\left(1+\frac{q}{M}+\frac{1}{M}\right)\log\left(M+q+1\right)-1-\left(\frac{q}{M}+\frac{1}{M}\right)\log(1+q)\right]+ \frac{S}{M}-\frac{\gamma}{g_c(M)}\log(1+q)\\
&=\frac{\gamma}{g_c(M)}\left[\left(1+\frac{q}{M}+\frac{1}{M}\right)\log\left(M+q+1\right)-1-\left(1+\frac{q}{M}+\frac{1}{M}\right)\log(1+q)\right]+ \frac{S}{M}\\
&=\frac{\gamma}{g_c(M)}\left(1+\frac{q}{M}+\frac{1}{M}\right)\log\left(\frac{M+q+1}{q+1}\right)+o(1)\stackrel{(a)}{=}o(1),
\end{aligned}
\end{equation}
where $(a)$ can be shown by considering two cases: (i) when $g_c(M)=\Theta(M)$, then since $q=\mathcal{O}(M)$ and $S$ is finite, $(a)$ is thus true; and (ii) when $g_c(M)=o(M)$, then we must have $q=\Theta(M)$ because $q=\frac{C_2Sg_c(M)}{\gamma}$ and $\frac{C_1Sg_c(M)}{\gamma}> M$ and $\frac{C_1}{C_2}=\Theta(1)$. Therefore $(a)$ is true. Combining results in (\ref{eq:Assumpt_Thm_1_I}) and (\ref{eq:Assumpt_Thm_1_II}), we show that $\log z_1-\log\nu\leq 1$ is true, and thus prove that the validity of the assumption in (\ref{eq:Thm1_Assump}) for the caching policy proposed in Theorem 1. This concludes the proof of Theorem 1.

\section{Proof of Proposition 1}
Before the proof of Proposition 1, we first state a Lemma:

{\em Lemma 3: (the original Lemma 1 in \cite{lee2019throughput}):} Denote $\displaystyle{\sum_{f=a}^b} (f+q)^{-\gamma}=H(a,b,\gamma,q)$. When $\gamma\neq 1$, we have
\begin{equation*}
\frac{1}{1-\gamma}\left[ (b+q+1)^{1-\gamma}-(a+q)^{1-\gamma}\right]\leq H(a,b,\gamma,q)\leq \frac{1}{1-\gamma}\left[ (b+q)^{1-\gamma}-(a+q)^{1-\gamma} \right] + (a+q)^{-\gamma}.
\end{equation*}

Suppose $g_c(M)=\frac{\rho M}{C_1S}$, where $\rho\geq \gamma$.  According to Theorem 1, we obtain $m^*=M$. Then observe that $z_f=\left(P_r(f)\right)^{\frac{1}{g_c(M)}}$. The outage probability is 
\begin{equation}\label{eq:App_Coro_2_0}
\begin{aligned}
&p_o=\sum_{f=1}^{M} P_r(f) e^{-g_c(M)\log\frac{z_f}{\nu}}=\sum_{f=1}^{M} P_r(f) \left(\frac{z_f}{\nu}\right)^{-g_c(M)}\\
&=\left(\nu\right)^{g_c(M)}\sum_{f=1}^{M} P_r(f) \left(\left(P_r(f)\right)^{\frac{1}{g_c(M)}}\right)^{-g_c(M)}=\left(\nu\right)^{g_c(M)}\sum_{f=1}^{M} P_r(f) \left(P_r(f)\right)^{-1}=\left(\nu\right)^{g_c(M)}M
\end{aligned}
\end{equation}
where
\begin{equation}\label{eq:App_Coro_2_1}
\left(\nu\right)^{g_c(M)}=\exp\left(\frac{\sum_{f=1}^{M}\log z_f - S}{M}\cdot g_c(M)\right)=e^{\frac{g_c(M)}{M}\sum_{f=1}^{M}\log z_f}\cdot e^{\frac{-Sg_c(M)}{M}}=e^{\frac{g_c(M)}{M}\sum_{f=1}^{M}\log z_f}\cdot e^{-\frac{\rho}{C_1}}.
\end{equation}
We then note that
\begin{equation}
\begin{aligned}\label{eq:App_Coro_2_5}
\sum_{f=1}^{M}\log z_f&=\sum_{f=1}^{M}\log (P_r(f))^{\frac{1}{g_c(M)}}=\frac{1}{g_c(M)}\sum_{f=1}^{M}\log P_r(f)=\frac{1}{g_c(M)}\sum_{f=1}^{M}\log \frac{(f+q)^{-\gamma}}{H(1,M,\gamma,q)}\\
&=\frac{-\gamma}{g_c(M)}\sum_{f=1}^{M}\log (f+q) - \frac{M}{g_c(M)}\log H(1,M,\gamma,q)\\
&\stackrel{(a)}{\leq}\frac{-\gamma}{g_c(M)}\left(\log(1+q)+(M+q)\log (M+q) - M -(1+q)\log(1+q) +1\right)\\
&\qquad-\frac{M}{g_c(M)}\log\left(\frac{1}{1-\gamma}\left((M+q+1)^{1-\gamma}-(1+q)^{1-\gamma}\right)\right),
\end{aligned}
\end{equation}
where $(a)$ is because
\begin{equation}\label{eq:App_Coro_2_2}
\sum_{f=1}^M \log (f+q) \geq \log(1+q)+(M+q)\log (M+q) - M -(1+q)\log(1+q) +1
\end{equation}
by Lemma 2 and 
\begin{equation}
H(1,M,\gamma,q)\geq \frac{1}{1-\gamma}\left((M+q+1)^{1-\gamma}-(1+q)^{1-\gamma}\right)
\end{equation}
by Lemma 3. It follows from (\ref{eq:App_Coro_2_0}), (\ref{eq:App_Coro_2_1}), and (\ref{eq:App_Coro_2_5}) that the outage probability can be upper bounded as:
\begin{equation}
\begin{aligned}\label{eq:App_Coro_2_3}
p_o&=\sum_{f=1}^{M} P_r(f) e^{-g_c(M)\log\frac{z_f}{\nu}}=M\left(\nu\right)^{g_c(M)}\\
&\leq Me^{\frac{g_c(M)}{M}\left[\frac{-\gamma}{g_c(M)}\left(\log(1+q)+(M+q)\log (M+q) - M -(1+q)\log(1+q) +1\right)\right]}\\
&\qquad\cdot e^{\frac{g_c(M)}{M}\left[-\frac{M}{g_c(M)}\log\left(\frac{1}{1-\gamma}\left((M+q+1)^{1-\gamma}-(1+q)^{1-\gamma}\right)\right)\right]}\cdot e^{-\frac{\rho}{C_1}}\\
&=M e^{-\frac{\rho}{C_1}}\cdot(1+q)^{\frac{-\gamma}{M}}\cdot (M+q)^{\frac{-\gamma}{M}(M+q)}\cdot e^{\gamma}\cdot(1+q)^{\frac{\gamma}{M}(1+q)}\cdot e^{\frac{-\gamma}{M}}\\
&\qquad\cdot \left[\frac{1}{1-\gamma}\left((M+q+1)^{1-\gamma}-(1+q)^{1-\gamma}\right)\right]^{-1} \\
\end{aligned}
\end{equation}
We let $D=\frac{q}{M}$. It follows from (\ref{eq:App_Coro_2_3}) that:
\begin{equation}
\begin{aligned}\label{eq:App_Coro_2_4}
&p_o=\sum_{f=1}^{M} P_r(f) e^{-g_c(M)\log\frac{z_f}{\nu}}\\
&\leq \frac{(1-\gamma) M e^{-\left(\frac{\rho}{C_1}-\gamma\right)}\cdot e^{\frac{-\gamma}{M}}}{(M+DM+1)^{1-\gamma}-(1+DM)^{1-\gamma}}\cdot(1+DM)^{\frac{-\gamma}{M}}\cdot (M+DM)^{\frac{-\gamma}{M}(M+DM)}\cdot(1+DM)^{\frac{\gamma}{M}(1+DM)}\\
&=\frac{(1-\gamma) M e^{-\left(\frac{\rho}{C_1}-\gamma\right)}\cdot e^{\frac{-\gamma}{M}}}{(M+DM+1)^{1-\gamma}-(1+DM)^{1-\gamma}}\cdot (M+DM)^{-\gamma(1+D)}\cdot(1+DM)^{\gamma D}\\
&=\frac{(1-\gamma) e^{-\left(\frac{\rho}{C_1}-\gamma\right)}\cdot e^{\frac{-\gamma}{M}}\cdot(1+D)^{-\gamma(1+D)}\cdot(D+\frac{1}{M})^{\gamma D}}{(1+D+\frac{1}{M})^{1-\gamma}-(D+\frac{1}{M})^{1-\gamma}}\cdot\frac{M\cdot M^{-\gamma(1+D)}\cdot M^{\gamma D}}{M^{1-\gamma}}\\
&=(1-\gamma) e^{-\left(\frac{\rho}{C_1}-\gamma\right)}\frac{(1+D)^{-\gamma(1+D)}\cdot(D)^{\gamma D}}{(1+D)^{1-\gamma}-(D)^{1-\gamma}}+o\left((1-\gamma) e^{-\left(\frac{\rho}{C_1}-\gamma\right)}\frac{(1+D)^{-\gamma(1+D)}\cdot(D)^{\gamma D}}{(1+D)^{1-\gamma}-(D)^{1-\gamma}}\right).
\end{aligned}
\end{equation}

\section{Proof of Theorem 2}

 When considering $g_c(M)=\frac{\rho M}{C_1S}$, by Proposition 1, we know that the outage probability of the cluster is upper bounded as
\begin{equation}\label{eq:Outage_Th_1}
p_o\leq(1-\gamma)e^{-(\frac{\rho}{C_1}-\gamma)}\frac{D^{\gamma D}(1+D)^{-\gamma(1+D)}}{(1+D)^{1-\gamma}-D^{1-\gamma}}.
\end{equation} 

To compute the througput of a cluster, we leverage the results in \cite{Franceschetti2007Closing}. Recall that when using the achievable scheme in Sec. III.A, the multi-hop approach proposed in \cite{Franceschetti2007Closing} is used for delivering both real and virtual files. We denote the throughput generated via transmitting real file as effective throughput; the throughput generated via transmitting virtual file as virtual throughput; and the sum of the real and virtual throughput as mixing throughput. Since only the effective throughput can be taken into account for $\overline{T}_{\text{user}}$, we want to compute its value.

To compute the effective throughput, our approach is to first compute the mixing throughput, and then exclude the virtual throughput from it. From the definition, we know:
\begin{equation}
\begin{aligned}\label{eq:Thm_Th_1}
\overline{T}_{\text{user}}&=\mathbb{E}_{\mathsf{n}, \mathsf{P}}\left[\min_{u\in\mathcal{U}}\mathbb{E}\left[C_{u}\cdot\mathsf{1}_{\mathsf{H}_u}\mid \mathsf{n}, \mathsf{P}\right]\right],
\end{aligned}
\end{equation}
where $C_{u}$ is the mixing throughput of user $u$; $\mathsf{1}_{\mathsf{H}_u}$ is the indicating function of the event $\mathsf{H}_u$ defined as $\mathsf{H}_u=\lbrace$the user $u$ can find the desired file in the cluster$\rbrace$. Thus, $\mathsf{1}_{\mathsf{H}_u}=1$ if user $u$ can find the desired file in the cluster; otherwise $\mathsf{1}_{\mathsf{H}_u}=0$. Then according to the result in \cite{Franceschetti2007Closing} and \cite{xue2006scaling} and due to the frequency reuse scheme among different clusters, we have the following Theorem:

{\em Theorem A.1 \cite{Franceschetti2007Closing,xue2006scaling}:} When using the proposed achievable scheme, with high probability (w.h.p.), users in a cluster with side length $\sqrt{\frac{g_c(M)}{N}}$ can achieve $C_{u}=\Omega\left(\frac{1}{K}\sqrt{\frac{1}{g_c(M)}}\right)$ of the mixing throughput simultaneously.

From Theorem A.1, we know that, w.h.p, there exists a $\epsilon=\Theta(1)>0$ such that $C_{u}\geq \frac{\epsilon}{K}\sqrt{\frac{1}{g_c(M)}}$ for all users. We note that both Theorem A.1 and event $\mathsf{1}_{\mathsf{H}_u}$ have the symmetry property for all users. It is then sufficient that we consider an arbitrary user in the network. We let $C_{\text{user}}=\frac{\epsilon}{K}\sqrt{\frac{1}{g_c(M)}}$ and define an event $\mathsf{H}=\lbrace$the user can find the desired file in the cluster$\rbrace$. Recall that $P_h=1-p_o$ is the file hit-rate. Then by using above arguments and (\ref{eq:Thm_Th_1}) and $g_c(M)=\frac{C_1S}{\rho M}$, we obtain:
\begin{equation}
\begin{aligned}\label{eq:Thm_Th_2}
\overline{T}_{\text{user}}&\geq\mathbb{E}_{\mathsf{n}, \mathsf{P}}\left[\mathbb{E}\left[C_{\text{user}}\cdot\mathsf{1}_{\mathsf{H}}\mid \mathsf{n}, \mathsf{P}\right]\right]=C_{\text{user}}\cdot\mathbb{E}_{\mathsf{n}, \mathsf{P}}\left[\mathbb{E}\left[\mathsf{1}_{\mathsf{H}}\mid \mathsf{n}, \mathsf{P}\right]\right]=C_{\text{user}}\cdot P_h\\
&=(1-p_o)C_{\text{user}}=\Omega\left(\frac{1-p_o}{K}\sqrt{\frac{1}{g_c(M)}}\right)=\Omega\left(\frac{1-p_o}{K}\sqrt{\frac{C_1 S}{\rho M}}\right).
\end{aligned}
\end{equation}
By combining (\ref{eq:Outage_Th_1}) and (\ref{eq:Thm_Th_2}), we finally obtain the asymptotic achievable throughput-outage tradeoff:
\begin{equation}
T(P_o)=\Omega\left(\frac{1-P_o}{K}\sqrt{\frac{C_1 S}{\rho M}}\right),P_o=(1-\gamma)e^{-(\frac{\rho}{C_1}-\gamma)}\frac{D^{\gamma D}(1+D)^{-\gamma(1+D)}}{(1+D)^{1-\gamma}-D^{1-\gamma}}.
\end{equation}

\section{Proof of Lemma 2}
By using the concept of Riemann sum in Calculus, we can obtain:
\begin{equation}
\begin{aligned}
&\sum_{f=1}^F \log(f+q)\leq \int_1^{F+1} \log (x+q)dx=(x+q)\log(x+q) -x \mid_1^{F+1} ;\\
&\sum_{f=1}^F \log(f+q)\geq \log(1+q)+\int_1^{F} \log (x+q)dx=\log(1+q)+\left[(x+q)\log(x+q) -x \mid_1^{F}\right].
\end{aligned}
\end{equation}
It follows that 
\begin{equation}
\begin{aligned}
&\sum_{f=1}^F \log(f+q)\leq (F+q+1)\log(F+q+1)-F-(1+q)\log(1+q);\\
&\sum_{f=1}^F \log(f+q)\geq \log(1+q)+(F+q)\log(F+q)-F-(1+q)\log(1+q)+1.
\end{aligned}
\end{equation}

\section{Proof of Proposition 2}
In the following, we will obtain both the upper and lower bounds of $p_o$ when $g_c(M)<\frac{\gamma M}{C_1S}$. As will be shown, the upper and lower bounds have the same expression. We can thus conclude the expression of $p_o$. When $g_c(M)<\frac{\gamma M}{C_1S}$, we have $m^*< M$ according to Theorem 1. Consequently, the outage probability is:
\begin{equation}\label{eq:App_Prop_2_0}
\begin{aligned}
&p_o=\sum_{f=1}^{M} P_r(f) e^{-g_c(M)\log\frac{z_f}{\nu}}=\sum_{f=1}^{m^*} P_r(f) \left(\frac{z_f}{\nu}\right)^{-g_c(M)}+\sum_{f=m^*+1}^M P_r(f)\\
&=\sum_{f=1}^{m^*} P_r(f)e^{-g_c(M)\log z_f}\cdot e^{g_c(M)\log\nu}+\sum_{f=m^*+1}^M P_r(f)=\sum_{f=1}^{m^*} P_r(f)(z_f)^{-g_c(M)}\cdot \nu^{g_c(M)}+\sum_{f=m^*+1}^M P_r(f)\\
&=\nu^{g_c(M)}\sum_{f=1}^{m^*} P_r(f)(P_r(f))^{\frac{-g_c(M)}{g_c(M)}}+\sum_{f=m^*+1}^M P_r(f)\\&
=m^*\nu^{g_c(M)}+\sum_{f=m^*+1}^M P_r(f)=1-\sum_{f=1}^{m^*} P_r(f)+m^*\nu^{g_c(M)}.
\end{aligned}
\end{equation}
To lower bound (\ref{eq:App_Prop_2_0}), we in the following upper bound $\sum_{f=1}^{m^*} P_r(f)$ and lower bound $m^*\nu^{g_c(M)}$, respectively.

We first derive the upper bound for $\sum_{f=1}^{m^*} P_r(f)$ as follows:
\begin{equation}
\begin{aligned}\label{eq:App_Prop_2_1}
&\sum_{f=1}^{m^*} P_r(f)=\frac{H(1,m^*,\gamma,q)}{H(1,M,\gamma,q)}\stackrel{(a)}{\leq} \frac{\frac{1}{1-\gamma}\left[(m^*+q)^{1-\gamma}-(q+1)^{1-\gamma}\right]+(q+1)^{-\gamma}}{\frac{1}{1-\gamma}\left[(M+q+1)^{1-\gamma}-(q+1)^{1-\gamma}\right]}\\
&=\frac{(m^*+q)^{1-\gamma}-(q+1)^{1-\gamma}+(1-\gamma)(q+1)^{-\gamma}}{(M+q+1)^{1-\gamma}-(q+1)^{1-\gamma}}=\frac{(m^*+q)^{1-\gamma}-(q+1)^{1-\gamma}}{(M+q)^{1-\gamma}-(q+1)^{1-\gamma}}+o\left(\xi\right),
\end{aligned}
\end{equation}
where $(a)$ is due to Lemma 3 and $\xi$ is with whatever order we have in (\ref{eq:App_Prop_2_1}). Thus, $o\left(\xi\right)$ here is simply to indicate some negligible terms that have even smaller order than the major term in (\ref{eq:App_Prop_2_1}). We use this for notation simplicity, and the same notation applies for all derivations in this Appendix.

We next derive the lower bound for $m^*\nu^{g_c(M)}$. Recall that $m^*=\frac{C_1Sg_c(M)}{\gamma}$. Then by using the same derivation as in (\ref{eq:App_Coro_2_1}), we obtain:
\begin{equation}\label{eq:App_Prop_2_2}
\begin{aligned}
m^*\nu^{g_c(M)}&=m^*\exp\left(\frac{\sum_{f=1}^{m^*}\log z_f - S}{m^*}\cdot g_c(M)\right)=m^*e^{\frac{g_c(M)}{m^*}\sum_{f=1}^{m^*}\log z_f}\cdot e^{\frac{-Sg_c(M)}{m^*}}\\
&= m^*e^{\frac{g_c(M)}{m^*}\sum_{f=1}^{m^*}\log z_f}\cdot e^{-\frac{\gamma}{C_1}}.
\end{aligned}
\end{equation}
To find the lower bound of (\ref{eq:App_Prop_2_2}), our approach is by obtaining the lower bound of $\sum_{f=1}^{m^*}\log z_f$. By following the same derivations as in (\ref{eq:App_Coro_2_5}), we obtain:
\begin{equation}\label{eq:App_Prop_2_3}
\sum_{f=1}^{m^*}\log z_f=\frac{-\gamma}{g_c(M)}\sum_{f=1}^{m^*}\log(f+q)-\frac{m^*}{g_c(M)}H(1,M,\gamma,q).
\end{equation}
Then by (\ref{eq:App_Prop_2_3}) and Lemmas 2 and 3, we obtain:
\begin{equation}
\begin{aligned}\label{eq:App_Prop_2_4}
\sum_{f=1}^{m^*}\log z_f&\geq\frac{-\gamma}{g_c(M)}\left((m^*+q+1)\log (m^*+q+1) - m^* -(1+q)\log(1+q)\right)\\
&\qquad-\frac{m^*}{g_c(M)}\log\left(\frac{1}{1-\gamma}\left((M+q)^{1-\gamma}-(1+q)^{1-\gamma}\right)+(1+q)^{-\gamma}\right).
\end{aligned}
\end{equation}
By substituting (\ref{eq:App_Prop_2_4}) into  (\ref{eq:App_Prop_2_2}), we obtain:
\begin{equation}
\begin{aligned}\label{eq:App_Prop_2_5}
&m^*\nu^{g_c(M)}\\
&\geq m^*e^{-\frac{\gamma}{C_1}}e^{\frac{g_c(M)}{m^*}\left[\frac{-\gamma}{g_c(M)}\left((m^*+q+1)\log (m^*+q+1) - m^* -(1+q)\log(1+q)\right)\right]}\\
&\qquad\cdot e^{\frac{g_c(M)}{m^*}\left[-\frac{m^*}{g_c(M)}\log\left(\frac{1}{1-\gamma}\left((M+q)^{1-\gamma}-(1+q)^{1-\gamma}\right)+(1+q)^{-\gamma}\right)\right]}\\
&=m^*e^{-\frac{\gamma}{C_1}}(m^*+q+1)^{\frac{-\gamma}{m^*}(m^*+q+1)}\cdot e^{\gamma}\cdot (1+q)^{\frac{\gamma}{m^*}(1+q)}\\
&\qquad\cdot\frac{1}{\frac{1}{1-\gamma}\left((M+q)^{1-\gamma}-(1+q)^{1-\gamma}\right)+(1+q)^{-\gamma}}\\
&=(1-\gamma)e^{-\gamma\left(\frac{1}{C_1}-1\right)}m^*\frac{(m^*+q+1)^{-\gamma}(m^*+q+1)^{\frac{-\gamma (q+1)}{m^*}}(1+q)^{\frac{\gamma (q+1)}{m^*}}}{(M+q)^{1-\gamma}-(1+q)^{1-\gamma}+(1-\gamma)(1+q)^{-\gamma}}\\
&=(1-\gamma)e^{-\gamma\left(\frac{1}{C_1}-1\right)}\left(\frac{m^*}{M}\right)^{1-\gamma}\frac{(1+\frac{q}{m^*}+\frac{1}{m^*})^{-\gamma}}{\left(1+\frac{q}{M}\right)^{1-\gamma}-\left(\frac{q}{M}+\frac{1}{M}\right)^{1-\gamma}+\frac{(1-\gamma)(1+q)^{-\gamma}}{M^{1-\gamma}}}\cdot \left(\frac{1+q}{m^*+q+1}\right)^{\frac{\gamma (q+1)}{m^*}}\\
&=(1-\gamma)e^{-\gamma\left(\frac{1}{C_1}-1\right)}\left(\frac{m^*}{M}\right)^{1-\gamma}\frac{(1+\frac{q}{m^*})^{-\gamma}}{\left(1+\frac{q}{M}\right)^{1-\gamma}-\left(\frac{q}{M}\right)^{1-\gamma}}\cdot \left(\frac{q}{m^*+q}\right)^{\frac{\gamma q}{m^*}}+o\left(\xi\right).
\end{aligned}
\end{equation}

Recall that $m^*=\frac{C_1Sg_c(M)}{\gamma}$ and $q=\frac{C_2Sg_c(M)}{\gamma}$. By using (\ref{eq:App_Prop_2_0}), (\ref{eq:App_Prop_2_1}), and (\ref{eq:App_Prop_2_5}), we obtain:
\begin{equation}
\begin{aligned}\label{eq:App_Prop_2_6}
&p_o\geq 1+(1-\gamma)e^{-\gamma\left(\frac{1}{C_1}-1\right)}\left(\frac{m^*}{M}\right)^{1-\gamma}\frac{(1+\frac{q}{m^*})^{-\gamma}}{\left(1+\frac{q}{M}\right)^{1-\gamma}-\left(\frac{q}{M}\right)^{1-\gamma}}\cdot \left(\frac{q}{m^*+q}\right)^{\frac{\gamma q}{m^*}}\\
&\qquad - \frac{(m^*+q)^{1-\gamma}-(q+1)^{1-\gamma}}{(M+q)^{1-\gamma}-(q+1)^{1-\gamma}}+o(\xi)\\
&=1+(1-\gamma)e^{-\gamma\left(\frac{1}{C_1}-1\right)}\left(\frac{C_1S}{\gamma}\frac{g_c(M)}{M}\right)^{1-\gamma}\frac{(1+\frac{C_2}{C_1})^{-\gamma}}{\left(1+\frac{C_2S}{\gamma}\frac{g_c(M)}{M}\right)^{1-\gamma}-\left(\frac{C_2S}{\gamma}\frac{g_c(M)}{M}\right)^{1-\gamma}}\cdot \left(\frac{C_2}{C_1+C_2}\right)^{\frac{\gamma C_2}{C_1}}\\
&\qquad - \left(\frac{m^*}{M}\right)^{1-\gamma}\frac{(1+\frac{q}{m^*})^{1-\gamma}-(\frac{q}{m^*}+\frac{1}{m^*})^{1-\gamma}}{(1+\frac{q}{M})^{1-\gamma}-(\frac{q}{M}+\frac{1}{M})^{1-\gamma}}+o(\xi)\\
&=1+(1-\gamma)e^{-\gamma\left(\frac{1}{C_1}-1\right)}\left(\frac{C_1S}{\gamma}\frac{g_c(M)}{M}\right)^{1-\gamma}\frac{\left(\frac{C_1}{C_1+C_2}\right)^{\gamma}\cdot\left(\frac{C_2}{C_1+C_2}\right)^{\gamma\frac{ C_2}{C_1}}}{\left(1+\frac{C_2S}{\gamma}\frac{g_c(M)}{M}\right)^{1-\gamma}-\left(\frac{C_2S}{\gamma}\frac{g_c(M)}{M}\right)^{1-\gamma}}\\
&\qquad - \left(\frac{C_1S}{\gamma}\frac{g_c(M)}{M}\right)^{1-\gamma}\frac{\left(1+\frac{C_2}{C_1}\right)^{1-\gamma}-\left(\frac{C_2}{C_1}\right)^{1-\gamma}}{\left(1+\frac{C_2S}{\gamma}\frac{g_c(M)}{M}\right)^{1-\gamma}-\left(\frac{C_2S}{\gamma}\frac{g_c(M)}{M}\right)^{1-\gamma}}+o\left(\xi\right).
\end{aligned}
\end{equation}

Similar to the above procedure, we derive the upper bound for $p_o$. Therefore, to obtain the upper bound of (\ref{eq:App_Prop_2_0}), we in the following obtain the lower bound of $\sum_{f=1}^{m^*} P_r(f)$ and upper bound of $m^*\nu^{g_c(M)}$, respectively. We first derive the lower bound for $\sum_{f=1}^{m^*} P_r(f)$ as follows:
\begin{equation}
\begin{aligned}\label{eq:App_Prop_3_1}
&\sum_{f=1}^{m^*} P_r(f)=\frac{H(1,m^*,\gamma,q)}{H(1,M,\gamma,q)}\stackrel{(a)}{\geq} \frac{\frac{1}{1-\gamma}\left[(m^*+q+1)^{1-\gamma}-(q+1)^{1-\gamma}\right]}{\frac{1}{1-\gamma}\left[(M+q)^{1-\gamma}-(q+1)^{1-\gamma}\right]+(q+1)^{-\gamma}}\\
&=\frac{(m^*+q+1)^{1-\gamma}-(q+1)^{1-\gamma}}{(M+q)^{1-\gamma}-(q+1)^{1-\gamma}+(1-\gamma)(q+1)^{-\gamma}}\\
&=\frac{(m^*+q)^{1-\gamma}-(q+1)^{1-\gamma}}{(M+q)^{1-\gamma}-(q+1)^{1-\gamma}}+o\left(\xi\right),
\end{aligned}
\end{equation}
where $(a)$ is due to Lemma 1 of \cite{lee2019throughput}.

We then derive the upper bound for $m^*\nu^{g_c(M)}$. Recall that $m^*=\frac{C_1Sg_c(M)}{\gamma}$ and by using the same derivation as in (\ref{eq:App_Prop_2_2}), we obtain:
\begin{equation}\label{eq:App_Prop_3_2}
\begin{aligned}
m^*\nu^{g_c(M)}= m^*e^{\frac{g_c(M)}{m^*}\sum_{f=1}^{m^*}\log z_f}\cdot e^{-\frac{\gamma}{C_1}}.
\end{aligned}
\end{equation}
Then by (\ref{eq:App_Prop_2_3}) and Lemmas 2 and 3, we obtain:
\begin{equation}
\begin{aligned}\label{eq:App_Prop_3_4}
\sum_{f=1}^{m^*}\log z_f&\leq\frac{-\gamma}{g_c(M)}\left(\log(1+q)+(m^*+q)\log (m^*+q) - m^* -(1+q)\log(1+q)+1\right)\\
&\qquad-\frac{m^*}{g_c(M)}\log\left(\frac{1}{1-\gamma}\left((M+q+1)^{1-\gamma}-(1+q)^{1-\gamma}\right)\right).
\end{aligned}
\end{equation}
By substituting (\ref{eq:App_Prop_3_4}) into  (\ref{eq:App_Prop_3_2}), we obtain:
\begin{equation}
\begin{aligned}\label{eq:App_Prop_3_5}
&m^*\nu^{g_c(M)}\\
&\leq m^*e^{-\frac{\gamma}{C_1}}e^{\frac{g_c(M)}{m^*}\left[\frac{-\gamma}{g_c(M)}\left(\log(1+q)+(m^*+q)\log (m^*+q) - m^* -(1+q)\log(1+q)+1\right)\right]}\\
&\qquad\cdot e^{\frac{g_c(M)}{m^*}\left[-\frac{m^*}{g_c(M)}\log\left(\frac{1}{1-\gamma}\left((M+q+1)^{1-\gamma}-(1+q)^{1-\gamma}\right)\right)\right]}\\
&=m^*e^{-\frac{\gamma}{C_1}}\cdot(1+q)^{\frac{-\gamma}{m^*}}\cdot(m^*+q)^{\frac{-\gamma}{m^*}(m^*+q)}\cdot e^{\gamma}\cdot (1+q)^{\frac{\gamma}{m^*}(1+q)}\cdot e^{\frac{-\gamma}{m^*}}\\
&\qquad\cdot \frac{1}{\frac{1}{1-\gamma}\left((M+q+1)^{1-\gamma}-(1+q)^{1-\gamma}\right)}\\
&=(1-\gamma)e^{-\gamma\left(\frac{1}{C_1}-1\right)}m^*\frac{(m^*+q)^{-\gamma}(m^*+q)^{\frac{-\gamma q}{m^*}}(1+q)^{\frac{\gamma q}{m^*}}e^{\frac{-\gamma}{m^*}}}{(M+q+1)^{1-\gamma}-(1+q)^{1-\gamma}}\\
&=(1-\gamma)e^{-\gamma\left(\frac{1}{C_1}-1\right)}\left(\frac{m^*}{M}\right)^{1-\gamma}\frac{(1+\frac{q}{m^*})^{-\gamma}\cdot e^{\frac{-\gamma}{m^*}}}{\left(1+\frac{q}{M}+\frac{1}{M}\right)^{1-\gamma}-\left(\frac{q}{M}+\frac{1}{M}\right)^{1-\gamma}}\cdot \left(\frac{1+q}{m^*+q}\right)^{\frac{\gamma (q+1)}{m^*}}\\
&=(1-\gamma)e^{-\gamma\left(\frac{1}{C_1}-1\right)}\left(\frac{m^*}{M}\right)^{1-\gamma}\frac{(1+\frac{q}{m^*})^{-\gamma}}{\left(1+\frac{q}{M}\right)^{1-\gamma}-\left(\frac{q}{M}\right)^{1-\gamma}}\cdot \left(\frac{q}{m^*+q}\right)^{\frac{\gamma q}{m^*}}+o\left(\xi\right).
\end{aligned}
\end{equation}

Recall that $m^*=\frac{C_1Sg_c(M)}{\gamma}$ and $q=\frac{C_2Sg_c(M)}{\gamma}$. Also observe that the final results of (\ref{eq:App_Prop_3_1}) and (\ref{eq:App_Prop_3_5}) are identical to (\ref{eq:App_Prop_2_1}) and (\ref{eq:App_Prop_2_5}), respectively. By using (\ref{eq:App_Prop_2_0}), (\ref{eq:App_Prop_3_1}), and (\ref{eq:App_Prop_3_5}), and following the similar derivations as in (\ref{eq:App_Prop_2_6}), we obtain:
\begin{equation}
\begin{aligned}\label{eq:App_Prop_3_6}
&p_o\leq 1+(1-\gamma)e^{-\gamma\left(\frac{1}{C_1}-1\right)}\left(\frac{C_1S}{\gamma}\frac{g_c(M)}{M}\right)^{1-\gamma}\frac{\left(\frac{C_1}{C_1+C_2}\right)^{\gamma}\cdot\left(\frac{C_2}{C_1+C_2}\right)^{\gamma\frac{ C_2}{C_1}}}{\left(1+\frac{C_2S}{\gamma}\frac{g_c(M)}{M}\right)^{1-\gamma}-\left(\frac{C_2S}{\gamma}\frac{g_c(M)}{M}\right)^{1-\gamma}}\\
&\qquad - \left(\frac{C_1S}{\gamma}\frac{g_c(M)}{M}\right)^{1-\gamma}\frac{\left(1+\frac{C_2}{C_1}\right)^{1-\gamma}-\left(\frac{C_2}{C_1}\right)^{1-\gamma}}{\left(1+\frac{C_2S}{\gamma}\frac{g_c(M)}{M}\right)^{1-\gamma}-\left(\frac{C_2S}{\gamma}\frac{g_c(M)}{M}\right)^{1-\gamma}}+o\left(\xi\right).
\end{aligned}
\end{equation}
Combining (\ref{eq:App_Prop_2_6}) and (\ref{eq:App_Prop_3_6}) completes the proof.

\section{Proof of Proposition 3}
From Proposition 2, when $\gamma>1$, $g_c(M)=o(M)$, and $q=o(M)$, we obtain:
\begin{equation}
\begin{aligned}\label{eq:App_Coro_4_0}
&p_o=1+(1-\gamma)e^{-\gamma\left(\frac{1}{C_1}-1\right)}\left(\frac{C_1S}{\gamma}\frac{g_c(M)}{M}\right)^{1-\gamma}\frac{\left(\frac{C_1}{C_1+C_2}\right)^{\gamma}\cdot\left(\frac{C_2}{C_1+C_2}\right)^{\gamma\frac{ C_2}{C_1}}}{\left(1+\frac{C_2S}{\gamma}\frac{g_c(M)}{M}\right)^{1-\gamma}-\left(\frac{C_2S}{\gamma}\frac{g_c(M)}{M}\right)^{1-\gamma}}\\
&\qquad - \left(\frac{C_1S}{\gamma}\frac{g_c(M)}{M}\right)^{1-\gamma}\frac{\left(1+\frac{C_2}{C_1}\right)^{1-\gamma}-\left(\frac{C_2}{C_1}\right)^{1-\gamma}}{\left(1+\frac{C_2S}{\gamma}\frac{g_c(M)}{M}\right)^{1-\gamma}-\left(\frac{C_2S}{\gamma}\frac{g_c(M)}{M}\right)^{1-\gamma}}\\
&= 1+(\gamma-1)e^{-\gamma\left(\frac{1}{C_1}-1\right)}\left(\frac{\gamma}{C_1S}\frac{M}{g_c(M)}\right)^{\gamma-1}\frac{\left(\frac{C_1}{C_1+C_2}\right)^{\gamma}\cdot\left(\frac{C_2}{C_1+C_2}\right)^{\gamma\frac{ C_2}{C_1}}}{\left(\frac{\gamma}{C_2S}\frac{M}{g_c(M)}\right)^{\gamma-1}-\left(\frac{\gamma M}{C_2Sg_c(M)+\gamma M}\right)^{\gamma-1}}\\
&\qquad - \left(\frac{\gamma}{C_1S}\frac{M}{g_c(M)}\right)^{\gamma-1}\frac{\left(\frac{C_1}{C_2}\right)^{\gamma-1}-\left(\frac{C_1}{C_1+C_2}\right)^{\gamma-1}}{\left(\frac{\gamma}{C_2S}\frac{M}{g_c(M)}\right)^{\gamma-1}-\left(\frac{\gamma M}{C_2Sg_c(M)+\gamma M}\right)^{\gamma-1}}\\
&=1+(\gamma-1)e^{-\gamma\left(\frac{1}{C_1}-1\right)}\left(\frac{\gamma}{C_1S}\frac{M}{g_c(M)}\right)^{\gamma-1}\frac{\left(\frac{C_1}{C_1+C_2}\right)^{\gamma}\cdot\left(\frac{C_2}{C_1+C_2}\right)^{\gamma\frac{ C_2}{C_1}}}{\left(\frac{\gamma}{C_2S}\frac{M}{g_c(M)}\right)^{\gamma-1}}\\
&\qquad - \left(\frac{\gamma}{C_1S}\frac{M}{g_c(M)}\right)^{\gamma-1}\frac{\left(\frac{C_1}{C_2}\right)^{\gamma-1}-\left(\frac{C_1}{C_1+C_2}\right)^{\gamma-1}}{\left(\frac{\gamma}{C_2S}\frac{M}{g_c(M)}\right)^{\gamma-1}}\\
&=1+(\gamma-1)e^{-\gamma\left(\frac{1}{C_1}-1\right)}\cdot\left(\frac{C_1}{C_1+C_2}\right)^{\gamma}\cdot\left(\frac{C_2}{C_1+C_2}\right)^{\gamma\frac{ C_2}{C_1}}\cdot\left(\frac{C_2}{C_1}\right)^{\gamma-1}\\
&\qquad-\left(\left(\frac{C_1}{C_2}\right)^{\gamma-1}-\left(\frac{C_1}{C_1+C_2}\right)^{\gamma-1}\right)\cdot\left(\frac{C_2}{C_1}\right)^{\gamma-1}
\end{aligned}
\end{equation}

\section{Proof of Corollary 3}
From Proposition 3, we know
\begin{equation}
\begin{aligned}
p_o=& 1+(\gamma-1)e^{-\gamma\left(\frac{1}{C_1}-1\right)}\cdot\left(\frac{C_1}{C_1+C_2}\right)^{\gamma}\cdot\left(\frac{C_2}{C_1+C_2}\right)^{\gamma\frac{ C_2}{C_1}}\cdot\left(\frac{C_2}{C_1}\right)^{\gamma-1}\\
&\qquad-\left(\left(\frac{C_1}{C_2}\right)^{\gamma-1}-\left(\frac{C_1}{C_1+C_2}\right)^{\gamma-1}\right)\cdot\left(\frac{C_2}{C_1}\right)^{\gamma-1}.
\end{aligned}
\end{equation}
Then when considering $g_c(M)=\frac{\alpha_1q}{S}$, we obtain $C_2=\frac{\gamma}{\alpha_1}$. Then when increasing $\alpha_1$ to a large number, this obtain $C_1$ very close to $1$ and $C_2$ very close to zero. As a result, we obtain
\begin{equation}
p_o=1+(\gamma-1)(1-\delta_1(\alpha_1))\cdot(1-\delta_2(\alpha_1))\cdot(1-\delta_3(\alpha_1))\cdot\delta_4(\alpha_1)-(1-\delta_5(\alpha_1)).
\end{equation}
Since $\delta_k(\alpha_1),k=1,...,5,$ can be arbitrarily small when increasing $\alpha_1$ to be sufficiently large, we conclude that $p_o=\epsilon(\alpha_1)$, where $\epsilon(\alpha_1)$ is arbitrarily small. Furthermore, if we let $\alpha_1\to\infty$, it follows that $\delta_4(\alpha_1)=\Theta\left(\frac{1}{(\alpha_1)^{\gamma-1}}\right)$ and $\delta_5=\Theta\left(\frac{1}{(\alpha_1)^{\gamma-1}}\right)$. As a result, we obtain
\begin{equation}
p_o=1+\Theta\left(\frac{1}{(\alpha_1)^{\gamma-1}}\right)-1+\Theta\left(\frac{1}{(\alpha_1)^{\gamma-1}}\right)=\Theta\left(\frac{1}{(\alpha_1)^{\gamma-1}}\right).
\end{equation}

\section{Proof of Theorem 3}

The procedure for proving Theorem 3 is same as for Theorem 2. We here consider $g_c(M)=o(M)$, $q=o(M)$, and $g_c(M)=\frac{\alpha_1q}{S}$, where $\alpha_1=\Omega(1)$. Consequently by Proposition 3, we obtain the outage probability:
\begin{equation}\label{eq:Outage_Th_3}
\begin{aligned}
p_o=&1+(\gamma-1)e^{-\gamma\left(\frac{1}{C_1}-1\right)}\cdot\left(\frac{C_1}{C_1+C_2}\right)^{\gamma}\cdot\left(\frac{C_2}{C_1+C_2}\right)^{\gamma\frac{ C_2}{C_1}}\cdot\left(\frac{C_2}{C_1}\right)^{\gamma-1}\\
&\qquad-\left(\left(\frac{C_1}{C_2}\right)^{\gamma-1}-\left(\frac{C_1}{C_1+C_2}\right)^{\gamma-1}\right)\cdot\left(\frac{C_2}{C_1}\right)^{\gamma-1}.
\end{aligned}
\end{equation} 

To derive the achievable $\overline{T}_{\text{user}}$, the same approach used for deriving Theorem 2 is adopted. Therefore, we obtain:
\begin{equation}
\begin{aligned}\label{eq:Thm3_Th_1}
\overline{T}_{\text{user}}&=\mathbb{E}\left[\min_{u\in\mathcal{U}}\mathbb{E}\left[C_{u}\cdot\mathsf{1}_{\mathsf{H}_u}\mid \mathsf{n}, \mathsf{P}\right]\right],
\end{aligned}
\end{equation}
Then again by exploiting Theorem A.1 (see Appendix D), we can use the same arguments as in Appendix D and $g_c(M)=\frac{\alpha_1q}{S}$, leading to:
\begin{equation}
\begin{aligned}\label{eq:Thm3_Th_2}
\overline{T}_{\text{user}}&\geq\mathbb{E}\left[\mathbb{E}\left[C_{\text{user}}\cdot\mathsf{1}_{\mathsf{H}}\mid \mathsf{n}, \mathsf{P}\right]\right]=C_{\text{user}}\cdot\mathbb{E}\left[\mathbb{E}\left[\mathsf{1}_{\mathsf{H}}\mid \mathsf{n}, \mathsf{P}\right]\right]=C_{\text{user}}\cdot P_h\\
&=(1-p_o)C_{\text{user}}=\Omega\left(\frac{1-p_o}{K}\sqrt{\frac{1}{g_c(M)}}\right)=\Omega\left(\frac{1-p_o}{K}\sqrt{\frac{S}{\alpha_1 q}}\right).
\end{aligned}
\end{equation}
By combining (\ref{eq:Outage_Th_3}) and (\ref{eq:Thm3_Th_2}), we finally obtain the asymptotic achievable throughput-outage performance:
\begin{equation}
\begin{aligned}
T(P_o)&=\Omega\left(\frac{(1-P_o)}{K}\sqrt{\frac{S}{\alpha_1 q}}\right),\\
P_o&=1+(\gamma-1)e^{-\gamma\left(\frac{1}{C_1}-1\right)}\cdot\left(\frac{C_1}{C_1+C_2}\right)^{\gamma}\cdot\left(\frac{C_2}{C_1+C_2}\right)^{\gamma\frac{ C_2}{C_1}}\cdot\left(\frac{C_2}{C_1}\right)^{\gamma-1}\\
&\qquad-\left(\left(\frac{C_1}{C_2}\right)^{\gamma-1}-\left(\frac{C_1}{C_1+C_2}\right)^{\gamma-1}\right)\cdot\left(\frac{C_2}{C_1}\right)^{\gamma-1}.
\end{aligned}
\end{equation}

\section{Proof of Theorem 4}

The procedure for the proof is as follows. We first consider the network having $\mathsf{n}=n=\omega(M)$ uniformly distributed users, and then derive the outer bound of $T_{\text{user}}(n)$ and $p_o(n)$, where $T_{\text{user}}(n)=\mathbb{E}_{\mathsf{P}\mid \mathsf{n}=n}\left[T_{\text{user}}(n,r)\right]$. Then, we compute the $\overline{T}_{\text{user}}$ and $p_o$ via accommodating different realizations of $\mathsf{n}$ with high probability.

Suppose the network has $\mathsf{n}=n=\omega(M)$ users, where the location placement $\mathsf{P}$ of users follows the BPP. We denote $\lambda(n,\mathsf{P})=\frac{\sum_{u\in \mathcal{U}}\overline{T}_u}{n}$ as the average throughput per user in the network and $\overline{L}(n,\mathsf{P})$ as the average distance between the source and destination in the network. By the definition in \cite{gupta2000capacity} and \cite{agarwal2004capacity}, we define the transport capacity as the number of {\em bit-meter} transported by the network per second, where one bit-meter is defined as one bit has been transported a distance of one meter toward its destination. Using Theorem 4.2 in \cite{agarwal2004capacity}, which describes the upper bound of the transport capacity of the network for any arbitrary placement of users and choice of transmission powers, we then obtain
\begin{equation}\label{eq:Theorem4_0}
\lambda(n,\mathsf{P})\overline{L}(n,\mathsf{P})n\leq \Theta\left(\sqrt{n}\right).
\end{equation}
Consequently, we obtain $\lambda(n,\mathsf{P})\leq \Theta\left(\frac{1}{\overline{L}(n,\mathsf{P})\sqrt{n}}\right)$. To compute the upper bound of $\lambda(n,\mathsf{P})$, we need to find $\overline{L}(n,\mathsf{P})$ as described below. First, we provide Lemmas 4 as follows.

{\em Lemma 4:} When $n=\omega(M)$ users are uniformly distributed within a network with unit size, the probability to have $N_{\text{D}}$ users within an area of size $A=o\left(\frac{N_{\text{D}}}{n}\right)$ is upper bounded by $o(1)$.
\begin{proof}
We denote $n_A$ as the number of users in an area with size $A$. Then according to Markov inequality, we can obtain
\begin{equation}\label{eq:Lemma3_1}
\mathbb{P}\left(n_A\geq N_{\text{D}}\right)\leq \frac{\mathbb{E}[n_A]}{N_{\text{D}}}=\frac{n\cdot A}{N_{\text{D}}}.
\end{equation}
Consequently, by letting $A=o\left(\frac{N_{\text{D}}}{n}\right)$, we complete the proof.
\end{proof}

We denote $n_{\text{s}}$ as the number of different users that a certain user searches through for obtaining the desired file and denote $p_{\text{miss}}(n)$ as the probability that this user cannot find the desired file from those users being searched. Then, we provide Lemma 5:

{\em Lemma 5:} Suppose $\gamma<1$. We then have the following results: when a user in the network searches through $n_{\text{s}}=o\left(\frac{M}{S}\right)$ different users, we obtain $p_{\text{miss}}(n_{\text{s}})\geq 1-o(1)$. Furthermore, when a user in the network searches through $n_{\text{s}}=\rho' M$ different users for some $\rho'$, we have the following results: (i) $p_{\text{miss}}(n_{\text{s}})\geq \epsilon_1'(\rho')$ if $\rho'=\Theta(1)$, where $\epsilon_1'(\rho')$ can be arbitrarily small as $\rho'$ is large enough; and (ii) $p_{\text{miss}}(n_{\text{s}})\geq (1-\gamma)e^{-(S\rho'-\gamma)}$ if $\rho'=\omega(1)$.
\begin{proof}
See Appendix N.
\end{proof}

From Lemmas 4 and 5, we conclude that to have a non-vanishing probability for a user to obtain the desired file (i.e., $p_{\text{miss}}(n)$ does not go to $1$), with high probability (w.h.p.), the distance between the source and destination is at least $\Theta\left(\sqrt{\frac{M}{Sn}}\right)$. As a result, $\overline{L}(n,\mathsf{P})= \Omega\left(\sqrt{\frac{M}{Sn}}\right)$. Furthermore, if we consider $\overline{L}(n,\mathsf{P})=\Theta\left(\sqrt{\frac{\rho' M}{Sn}}\right)$, we know that, w.h.p., the distance between a source-destination pair is $\mathcal{O}\left(\sqrt{\frac{\rho' M}{Sn}}\right)$; otherwise we should have $\overline{L}(n,\mathsf{P})= \omega\left(\sqrt{\frac{\rho' M}{Sn}}\right)$. As a result, w.h.p., the number of users searched by a user is $n_{\text{s}}=\mathcal{O}\left(\frac{\rho' M}{S}\right)$. Recall that $p_{\text{miss}}(n_{\text{s}})$ characterizes the outage probability when $n_{\text{s}}$ users are searched. Also, note that above arguments apply to any number of users $n=\omega(M)$ and any $\mathsf{P}$ of the network. Accordingly, by combining this with Lemma 5, it follows that
\begin{equation}
\lambda(n,\mathsf{P})=\mathcal{O}\left(\sqrt{\frac{S}{\rho'M}}\right)
\end{equation}
with $p_o(n)\geq  \epsilon_1'(\rho')$ if $\rho'=\Theta(1)$, where $\epsilon_1'(\rho')$ can be arbitrarily small; and $p_o(n)\geq (1-\gamma)e^{-(\rho'-\gamma)}$ if $\rho'=\omega(1)$. Then since $\lambda(n,\mathsf{P})=\frac{\sum_{u\in \mathcal{U}}\overline{T}_u}{n}\geq T_{\text{user}}(n,\mathsf{P})$, we conclude that for any $n=\omega(M)$, we must have
\begin{equation}\label{eq:Theorem4_1}
\begin{aligned}
&T_{\text{user}}(n)=\mathcal{O}\left(\sqrt{\frac{S}{\rho'M}}\right),p_o(n)\geq  \epsilon_1'(\rho'),\text{ if }\rho'=\Theta(1);\\
&T_{\text{user}}(n)=\mathcal{O}\left(\sqrt{\frac{S}{\rho'M}}\right),p_o(n)\geq (1-\gamma)e^{-(\rho'-\gamma)},\text{ if }\rho'=\omega(1).
\end{aligned}
\end{equation}
To complete the proof of Theorem 4, we provide Lemma 6:

{\em Lemma 6:} Let $N$ be the density of a Poisson distribution in which $\mathbb{P}_N(n)=\frac{N^n}{n!}e^{-N}$. Then suppose $U=o(N)$, the following can be obtained:
\begin{equation}
\sum_{n=0}^{U}\mathbb{P}_N(n)\leq o(1).
\end{equation}
\begin{proof}
We denote $X$ as the Poisson random variable with density $N$. According to Chernoff bound, we obtain
\begin{equation}
\sum_{n=0}^{U}\mathbb{P}_N(n)=\mathbb{P}(X\leq U)=\mathbb{P}(e^{-tX}\geq e^{-tU}) \leq \frac{\mathbb{E}[e^{-tX}]}{e^{-tU}}=e^{tU}\mathbb{E}[e^{-tX}],
\end{equation}
where $t>0$. Then we observe that the moment generating function of $X$ gives $\mathbb{E}[e^{-tX}]=e^{N(e^{-t}-1)}$. By letting $t=1$, it follows that
\begin{equation}
\sum_{n=0}^{U}\mathbb{P}_N(n)\leq e^{U}\cdot e^{-N(1-\frac{1}{e})}=e^{-(N(1-\frac{1}{e})-U)}.
\end{equation}
By letting $U=o(N)$, we obtain $(N(1-\frac{1}{e})-U)=\Theta(N)\to\infty$. This leads to
\begin{equation}
\sum_{n=0}^{U}\mathbb{P}_N(n)\leq o(1).
\end{equation}
\end{proof}

Finally, recall that we consider $M=o(N)$ when $\gamma<1$. Consequently, for the adopted network, we have $\mathbb{P}_N\left(n=\omega(M)\right)=1-o(1)$ according to Lemma 6. In other words, w.h.p., we have $n=\omega(M)$. It follows from (\ref{eq:Theorem4_1}) and Lemma 6 that
\begin{equation}
\begin{aligned}
&\overline{T}_{\text{user}}=\mathcal{O}\left(\sqrt{\frac{S}{\rho'M}}\right),p_o\geq  \epsilon_1'(\rho'),\text{ if }\rho'=\Theta(1);\\
&\overline{T}_{\text{user}}=\mathcal{O}\left(\sqrt{\frac{S}{\rho'M}}\right),p_o\geq (1-\gamma)e^{-(\rho'-\gamma)},\text{ if }\rho'=\omega(1).
\end{aligned}
\end{equation}
This leads to Theorem 4.

\section{Proof of Theorem 5}

The procedure for the proof of Theorem 5 is similar to the proof for Theorem 4. Here, we again first derive the upper bound of $\lambda(n,\mathsf{P})$ by computing $\overline{L}(n,\mathsf{P})$. To do this, we first provide Lemmas 7 and 8 that will be used later:

{\em Lemma 7:} When $n=\omega(q)$ users are uniformly distributed within a network with unit size, the minimum size of an area to have $\Theta\left(\frac{q}{S}\right)$ users with high probability is $\Theta\left(\frac{q}{Sn}\right)$.
\begin{proof}
This is proved by using Lemma 4.
\end{proof}

{\em Lemma 8:} Suppose $\gamma>1$ and $n=\omega(q)$. Considering $q=o(M)$, we have the following results: (i) when a user searches through $n_{\text{s}}=o\left(\frac{q}{S}\right)$ different users in the network, we obtain $p_{\text{miss}}(n)\geq 1-o(1)$; (ii) when a user searches through $n_{\text{s}}=\frac{\alpha_1' q}{S}$ different users, where $\alpha_1'=\Theta(1)>0$, we obtain $p_{\text{miss}}(n)\geq \epsilon_{\text{miss}}(\alpha_1')$, where $\epsilon_{\text{miss}}(\alpha_1')=\Theta(1)>0$ can be arbitrarily small; (iii) when a user searches through $n_{\text{s}}=\frac{\alpha_1' q}{S}<\frac{M}{S}$ different users, where $\alpha_1'=\mathcal{O}\left(q^{\frac{1}{\gamma-1}}\right)\to\infty$, we obtain: $p_{\text{miss}}(n)\geq\Theta\left(\frac{1}{(\alpha_1')^{\gamma-1}}\right)$.
\begin{proof}
When letting $n_{\text{s}}=o\left(\frac{q}{S}\right)$, we obtain:
\begin{equation}
\begin{aligned}\label{eq:Lemma_8_2}
p_{\text{miss}}(n)&\geq 1-\sum_{f=1}^{Sn_{\text{s}}}P_r(f)= 1-\frac{H(1,Sn_{\text{s}},\gamma,q)}{H(1,M,\gamma,q)}\stackrel{(a)}{\geq}1 - \frac{\frac{1}{1-\gamma}\left[(Sn_{\text{s}}+q)^{1-\gamma}-(1+q)^{1-\gamma}\right]+(1+q)^{-\gamma}}{\frac{1}{1-\gamma}\left[(M+q+1)^{1-\gamma}-(1+q)^{1-\gamma}\right]}\\
&\geq 1 - \frac{\frac{1}{1-\gamma}\left[(Sn_{\text{s}}+q)^{1-\gamma}-(1+q)^{1-\gamma}\right]}{\frac{1}{1-\gamma}\left[(M+q+1)^{1-\gamma}-(1+q)^{1-\gamma}\right]}-o(1)=1-\frac{(1+q)^{1-\gamma}-(Sn_{\text{s}}+q)^{1-\gamma}}{(1+q)^{1-\gamma}-(M+q)^{1-\gamma}}-o(1)\\
&=1-\frac{1-\left(\frac{1+q}{Sn_{\text{s}}+q}\right)^{\gamma-1}}{1-\left(\frac{1+q}{M+q}\right)^{\gamma-1}}-o(1)\stackrel{(b)}{=}1-o(1),
\end{aligned}
\end{equation}
where $(a)$ is due to Lemma 3 and $(b)$ is due to $n_{\text{s}}=o\left(\frac{q}{S}\right)$. When letting $n_{\text{s}}=\frac{\alpha_1'q}{S}$, from (\ref{eq:Lemma_8_2}), we obtain:
\begin{equation}
\begin{aligned}\label{eq:Lemma_8_1}
p_{\text{miss}}(n)&\geq 1-\frac{1-\left(\frac{1+q}{Sn_{\text{s}}+q}\right)^{\gamma-1}}{1-\left(\frac{1+q}{M+q}\right)^{\gamma-1}}=\frac{\left(\frac{1+q}{Sn_{\text{s}}+q}\right)^{\gamma-1}-\left(\frac{1+q}{M+q}\right)^{\gamma-1}}{1-\left(\frac{1+q}{M+q}\right)^{\gamma-1}}\geq \left(\frac{1+q}{Sn_{\text{s}}+q}\right)^{\gamma-1}-\left(\frac{1+q}{M+q}\right)^{\gamma-1}\\
&\left(\frac{1+q}{Sn_{\text{s}}+q}\right)^{\gamma-1} - \Theta\left(\frac{1}{M^{\gamma-1}}\right) \stackrel{(a)}{=}\epsilon_{\text{miss}}(\alpha_1')-o(1),
\end{aligned}
\end{equation}
where $(a)$ is due to $\alpha_1'=\Theta(1)>0$. Finally, by using the similar derivations in (\ref{eq:Lemma_8_2}) and (\ref{eq:Lemma_8_1}) and letting $n_{\text{s}}=\frac{\alpha_1' q}{S}<\frac{M}{S}$, where $\alpha_1'=\mathcal{O}\left(q^{\frac{1}{\gamma-1}}\right)\to\infty$, we obtain:
\begin{equation}
\begin{aligned}\label{eq:Lemma_8_3}
p_{\text{miss}}(n)&\geq 1 - \frac{\frac{1}{1-\gamma}\left[(Sn_{\text{s}}+q)^{1-\gamma}-(1+q)^{1-\gamma}\right]+(1+q)^{-\gamma}}{\frac{1}{1-\gamma}\left[(M+q+1)^{1-\gamma}-(1+q)^{1-\gamma}\right]}\geq \left(\frac{1+q}{Sn_{\text{s}}+q}\right)^{\gamma-1}-\Theta\left(\frac{1}{1+q}\right)\\ &
=\left(\frac{1+q}{\alpha_1' q+q}\right)^{\gamma-1}-\Theta\left(\frac{1}{1+q}\right)=\left(\frac{\frac{1}{q}+1}{\alpha_1'+1}\right)^{\gamma-1}-\Theta\left(\frac{1}{1+q}\right)=\Theta\left(\frac{1}{(\alpha_1')^{\gamma-1}}\right).
\end{aligned}
\end{equation}
\end{proof}

From Lemmas 7 and 8, we conclude that to have a non-vanishing probability for a user to obtain the desired file, w.h.p., the distance between the source and destination is at least $\Theta\left(\sqrt{\frac{q}{Sn}}\right)$. Furthermore, if we consider $\overline{L}(n,\mathsf{P})=\Theta\left(\sqrt{\frac{\alpha_1' q}{Sn}}\right)$, we know that (w.h.p.) the distance between a source-destination pair is $\mathcal{O}\left(\sqrt{\frac{\alpha_1' q}{Sn}}\right)$; otherwise we should have $\overline{L}(n,\mathsf{P})= \omega\left(\sqrt{\frac{\alpha_1'q}{Sn}}\right)$. As a result, w.h.p., the number of users searched by a user is $n_s=\mathcal{O}\left(\frac{\alpha_1' q}{S}\right)$. Note that above arguments are valid for any $n=\omega(q)$ and $\mathsf{P}$. Consequently, by combining this with Lemma 8 and again using $T_{\text{user}}(n,\mathsf{P})\leq  \lambda(n,\mathsf{P})=\mathcal{O}\left(\frac{1}{\overline{L}(n,\mathsf{P})\sqrt{n}}\right)$, we conclude that for any $n=\omega(q)$, we must have
\begin{equation}\label{eq:Theorem5_1}
T_{\text{user}}(n)=\mathcal{O}\left(\sqrt{\frac{S}{\alpha_1' q}}\right)
\end{equation}
with $p_o(n)\geq \epsilon_{\text{miss}}(\alpha_1')$ when $\alpha_1'=
\Theta(1)$; and $p_o(n)\geq\Theta\left(\frac{1}{(\alpha_1')^{\gamma-1}}\right)$ when $\alpha_1'=\omega(1)$ and $\alpha_1'=o\left(q^{\frac{1}{\gamma-1}}\right)$. Finally, recall that we consider $q=o(N)$ when $\gamma>1$. Consequently, according to (\ref{eq:Theorem5_1}) and Lemma 6, we conclude:
\begin{equation}
\begin{aligned}
\overline{T}_{\text{user}}=\mathcal{O}\left(\sqrt{\frac{S}{\alpha_1' q}}\right),p_o\geq\epsilon_2'(\alpha_1'),
\end{aligned}
\end{equation}
in which $\epsilon_2'(\alpha_1')=\Theta(1)>0$ when $\alpha_1'=\Theta(1)$, where $\epsilon_2'(\alpha_1')$ can be arbitrarily small; $\epsilon_2'(\alpha_1')=\Theta\left(\frac{1}{(\alpha_1')^{\gamma-1}}\right)$ when $\alpha_1'=\omega(1)$ and $\alpha_1'=o\left(q^{\frac{1}{\gamma-1}}\right)$.

\section{Proof of Theorem 8}

We consider $m^*=o(M)$ and $m^*=\frac{Sg_c(M)}{\gamma}\to\infty$ when $g_c(M)\to\infty$. From (\ref{eq:App_Prop_2_0}), we know 
\begin{equation}\label{eq:Thm8_0}
p_o=1-\sum_{f=1}^{m^*} P_r(f)+m^*\nu^{g_c(M)}.
\end{equation}
By using the derivations in (\ref{eq:App_Prop_2_5}) and (\ref{eq:App_Prop_3_5}) and considering $\gamma>1$ and $q=\Theta(1)$, we obtain:
\begin{equation}\label{eq:Thm8_1}
\Theta\left(\frac{1}{(m^*)^{\gamma-1}}\right)\leq m^*\nu^{g_c(M)}\leq \Theta\left(\frac{1}{(m^*)^{\gamma-1}}\right).
\end{equation}
Then notice that when $\gamma>1$, $q=\Theta(1)$, and $g_c(M)\to\infty$, we have
\begin{equation}\label{eq:Thm8_2}
\sum_{f=1}^{m^*} P_r(f)=\sum_{f=1}^{m^*} \frac{(f+q)^{-\gamma}}{\sum_{m=1}^{M} (m+q)^{-\gamma}}=\frac{\sum_{f=1}^{m^*} (f+q)^{-\gamma}}{\sum_{f=1}^{M} (f+q)^{-\gamma}}=\frac{\sum_{f=q+1}^{m^*+q+1}f^{-\gamma}}{\sum_{f=q+1}^{M+q+1}f^{-\gamma}}=\frac{\sum_{f=1}^{m^*+q+1} f^{-\gamma}-\sum_{f=1}^{q} f^{-\gamma}}{\sum_{f=1}^{M+q+1} f^{-\gamma}-\sum_{f=1}^{q} f^{-\gamma}}.
\end{equation}
Observe that $\sum_{f=1}^{\infty}f^{-\gamma}$ converges when $\gamma>1$. We let $\sum_{f=1}^{M+q+1}f^{-\gamma}=\zeta$. It follows from (\ref{eq:Thm8_2}) and Lemma 3 that 
\begin{equation}\label{eq:Thm8_3}
\begin{aligned}
&\sum_{f=1}^{m^*} P_r(f)=\frac{\zeta-\sum_{f=m^*+q+2}^M f^{-\gamma}-\sum_{f=1}^{q} f^{-\gamma}}{\zeta-\sum_{f=1}^{q} f^{-\gamma}}=1-\frac{\sum_{f=m^*+q+2}^M f^{-\gamma}}{\zeta-\sum_{f=1}^{q} f^{-\gamma}}\\
&\geq 1-\frac{\frac{1}{1-\gamma}\left[M^{1-\gamma}-(m^*+q+2)^{1-\gamma}\right]+(m^*+q+2)^{-\gamma}}{\zeta-\sum_{f=1}^{q} f^{-\gamma}}\\
&=1-\frac{\left[(m^*+q+2)^{1-\gamma}-M^{1-\gamma}\right]+(\gamma-1)(m^*+q+2)^{-\gamma}}{(\gamma-1)(\zeta-\sum_{f=1}^{q} f^{-\gamma})}=1-\Theta\left(\frac{1}{(m^*)^{\gamma-1}}\right).
\end{aligned}
\end{equation}
Consequently, by using (\ref{eq:Thm8_0}), (\ref{eq:Thm8_1}), and (\ref{eq:Thm8_3}), we obtain $p_o\leq \Theta\left(\frac{1}{(Sg_c(M))^{\gamma-1}}\right)$.

To compute $\overline{T}_{\text{user}}$, we directly exploit the derivations in (\ref{eq:Thm3_Th_2}), leading to
\begin{equation}
\overline{T}_{\text{user}}=\Omega\left(\frac{1-p_o}{K}\sqrt{\frac{1}{g_c(M)}}\right).
\end{equation}
Then, since $p_o\leq \Theta\left(\frac{1}{(Sg_c(M))^{\gamma-1}}\right)$, we can let $g_c(M)=\frac{\alpha_{1,\text{zip}}'}{S}\to\infty$, where $\alpha_{1,\text{zip}}'$ is an arbitrary function such that $\alpha_{1,\text{zip}}'\to\infty$ when $M\to\infty$, and obtain the achievable throughput-outage performance:
\begin{equation}
T(P_o)=\Omega\left(\sqrt{\frac{S}{\alpha_{1,\text{zip}}'}}\right),P_o=\Theta\left(\frac{1}{(\alpha_{1,\text{zip}}')^{\gamma-1}}\right)=o(1).
\end{equation}

\section{Proof of Theorem 10}

The proof is identical to Theorem 5 with minor modifications. We first provide Lemma 9:

{\em Lemma 9:} Suppose $\gamma>1$, $n=\omega(q)$, and $q=\Theta(1)$. Let $\alpha_{2,\text{zip}}'=\Omega(1)$. When $n_{\text{s}}=\Theta\left(\frac{\alpha_{2,\text{zip}}' q}{S}\right)$, we have $p_{\text{miss}}(n)\geq \Theta\left(\frac{1}{(\alpha_{2,\text{zip}}')^{\gamma-1}}\right)$.
\begin{proof}
 By using (\ref{eq:Thm8_2}), (\ref{eq:Thm8_3}), Lemma 3, and $q=\Theta(1)$, we then know:
 \begin{equation}\label{eq:Thm10_1}
\begin{aligned}
&p_{\text{miss}}(n)=1-\sum_{f=1}^{Sn_{\text{s}}} P_r(f)=\frac{\sum_{f=Sn_{\text{s}}+q+2}^M f^{-\gamma}}{\zeta-\sum_{f=1}^{q} f^{-\gamma}}\geq \frac{\frac{1}{1-\gamma}\left[(M+1)^{1-\gamma}-(Sn_{\text{s}}+q+2)^{1-\gamma}\right]}{\zeta-\sum_{f=1}^{q} f^{-\gamma}}\\
&=\frac{\left[(Sn_{\text{s}}+q+2)^{1-\gamma}-(M+1)^{1-\gamma}\right]}{(\gamma-1)(\zeta-\sum_{f=1}^{q} f^{-\gamma})}=\Theta\left(\frac{1}{(\alpha_{2,\text{zip}}' q)^{\gamma-1}}\right)=\Theta\left(\frac{1}{(\alpha_{2,\text{zip}}')^{\gamma-1}}\right).
\end{aligned}
\end{equation}
\end{proof}

Then following similar derivations and arguments in Theorem 5, we know when if $\overline{L}(n)=\Theta\left(\frac{\alpha_{2,\text{zip}}' q}{Sn}\right)$, w.h.p., $n_{\text{s}}=\mathcal{O}\left(\frac{\alpha_{2,\text{zip}}' q}{S}\right)$ for any user in the network. Therefore, by using Lemma 9 and the same approach as that in Theorem 5, we obtain that the throughput-outage performance of the network is dominated by:
\begin{equation}
\begin{aligned}
&T(P_o)=\Theta\left(\sqrt{\frac{S}{\alpha_{2,\text{zip}}'}}\right),P_o=\Theta\left(\frac{1}{(\alpha_{2,\text{zip}}')^{\gamma-1}}\right).
\end{aligned}
\end{equation}

\section{Proof of Lemma 5}

The proof of the case that $n_{\text{s}}=o\left(\frac{M}{S}\right)$ is simple. In this case, the best $p_{\text{miss}}(n_{\text{s}})$ happens when all users being visited cache different files. Accordingly, we have
\begin{equation}
\begin{aligned}\label{eq:Lemma_5_1}
p_{\text{miss}}(n_{\text{s}})&\geq 1-\sum_{f=1}^{Sn_{\text{s}}}P_r(f)= 1-\frac{H(1,Sn_{\text{s}},\gamma,q)}{H(1,M,\gamma,q)}\stackrel{(a)}{\geq} 1 - \frac{\frac{1}{1-\gamma}\left[(Sn_{\text{s}}+q)^{1-\gamma}-(1+q)^{1-\gamma}\right]+(1+q)^{-\gamma}}{\frac{1}{1-\gamma}\left[(M+q+1)^{1-\gamma}-(1+q)^{1-\gamma}\right]}\\
&=1 - \frac{\frac{1}{1-\gamma}\left[(Sn_{\text{s}}+q)^{1-\gamma}-(1+q)^{1-\gamma}\right]}{\frac{1}{1-\gamma}\left[(M+q+1)^{1-\gamma}-(1+q)^{1-\gamma}\right]}-o(1)\stackrel{(b)}{=}1-o(1),
\end{aligned}
\end{equation}
where $(a)$ is due to Lemma 3; and $(b)$ is because 
\begin{equation}
\frac{\frac{1}{1-\gamma}\left[(Sn_{\text{s}}+q)^{1-\gamma}-(1+q)^{1-\gamma}\right]}{\frac{1}{1-\gamma}\left[(M+q+1)^{1-\gamma}-(1+q)^{1-\gamma}\right]}=o(1)
\end{equation}
for either $q=o(M)$ or $q=\Theta(M)$.

For the other cases, we recall that we adopt the decentralized random policy $P_c(\cdot)$. Accordingly, when considering the case that $n_{\text{s}}=\rho' M$, the probability that a user cannot find its desired file, i.e., $p_{\text{miss}}(n_{\text{s}})$, can be described as \cite{Ji:Th_Out_toff,lee2019throughput}:
\begin{equation}
p_{\text{miss}}(n_{\text{s}})=\sum_{f=1}^M P_r(f)(1-P_c(f))^{n_{\text{s}}}.
\end{equation}
Accordingly, the minimum of $p_{\text{miss}}(n_{\text{s}})$ can be obtained by optimizing the caching policy using:
\begin{equation}
\begin{aligned}\label{eq:Lemma5_Opt}
\min &\quad\sum_{f=1}^M P_r(f)(1-P_c(f))^{n_{\text{s}}}\\
 s.t. &\quad\sum_{f=1}^M P_c(f) = S\\
 & 0\leq P_c(f)\leq 1,\forall f=1,2,...,M.
\end{aligned}
\end{equation}
Observe that (\ref{eq:Lemma5_Opt}) has a very similar optimization problem to that being optimized in Theorem 1 of \cite{lee2019throughput} except that here $\sum_{f=1}^M P_c(f)$ is equal to $S$ instead of $1$ and that constraints $P_c(f)\leq 1,\forall f,$ need to be satisfied. To accommodate the additional constraints that $P_c(f)\leq 1,\forall f$, we adopt the same approach as that for proving Theorem 1 of this paper -- we assume that these constraints are satisfied when finding the solution, and then show that the assumption is indeed true for the obtained solution. Accordingly, we denote $P_{c,n_{\text{s}}}^{**}(f)$ as the optimum solution of (\ref{eq:Lemma5_Opt}); denote $m^{**}_{n_{\text{s}}}$ as the smallest index such that $P_c(m^{**}_{n_{\text{s}}}+1)=0$; and let $C^{**}_{2,n_{\text{s}}}=\frac{\gamma q}{n_{\text{s}}-1}$. Then, by following the similar procedure as that for proving Theorem 1 of \cite{lee2019throughput} (see details in Appendix A in \cite{lee2019throughput}), we can obtain:
\begin{equation}\label{eq:Lemma_5_Opt_Cach}
P_{c,n_{\text{s}}}^{**}(f)=\left[1-\frac{\nu^{**}_{n_{\text{s}}}}{z_{f,n_{\text{s}}}^{**}}\right]^+,
\end{equation}
where $\nu^{**}_{n_{\text{s}}}=\frac{m_{n_{\text{s}}}^{**}-S}{\sum_{f=1}^M \frac{1}{z_{f,n_{\text{s}}}^{**}}}$; $z_{f,n_{\text{s}}}^{**}=\left(P_r(f)\right)^{\frac{1}{n_{\text{s}}-1}}$; and
\begin{equation}
m_{n_{\text{s}}}^{**}=\Theta\left(\min\left(\frac{C^{**}_{1,n_{\text{s}}}n_{\text{s}}}{\gamma}\right),M\right),
\end{equation}
where $C^{**}_{1,n_{\text{s}}}\geq S$ is the solution of the equality $C^{**}_{1,n_{\text{s}}}=S+C^{**}_{2,n_{\text{s}}}\log \left(1+\frac{C^{**}_{1,n_{\text{s}}}}{C^{**}_{2,n_{\text{s}}}}\right)$. It should be noticed that now $\sum_{f=1}^M P_c(f)$ is equal to $S$ instead of $1$ when repeating the similar procedure as that in Appendix in \cite{lee2019throughput}. Finally, we can observe that $z_{f,n_{\text{s}}}^{**}=\left(P_r(f)\right)^{\frac{1}{n_{\text{s}}-1}}>0$, and thus $\nu^{**}_{n_{\text{s}}}=\frac{m_{n_{\text{s}}}^{**}-S}{\sum_{f=1}^M \frac{1}{z_{f,n_{\text{s}}}^{**}}}>0$ as well. Accordingly, we observe from (\ref{eq:Lemma_5_Opt_Cach}) that the assumption that $P_{c,n_{\text{s}}}^{**}(f)\leq 1,\forall f,$ is true.

Based on the derived caching policy $P_{c,n_{\text{s}}}^{**}(\cdot)$, we now derive $p_{\text{miss}}(n_{\text{s}})$. Suppose we let $\rho'$ be large enough such that $\frac{C^{**}_{1,n_{\text{s}}}n_{\text{s}}}{\gamma}\geq M$. Note that this is always possible as $\rho'=\Omega(1)$. We denote $\phi=n_{\text{s}}-1$. Then, noticing that $\nu^{**}_{n_{\text{s}}}=\frac{M-S}{\sum_{f=1}^M \frac{1}{z_{f,n_{\text{s}}}^{**}}}$, we obtain:
\begin{equation}
\begin{aligned}\label{eq:Lemma5_1}
&p_{\text{miss}}(n_{\text{s}})=\sum_{f=1}^M P_r(f)(1-P_c(f))^{n_{\text{s}}}=\sum_{f=1}^M P_r(f)\left(\frac{\nu^{**}_{n_{\text{s}}}}{z_{f,n_{\text{s}}}^{**}}\right)^{n_{\text{s}}}=(\nu^{**}_{n_{\text{s}}})^{n_{\text{s}}}\sum_{f=1}^M \frac{P_r(f)}{(z_{f,n_{\text{s}}}^{**})^{n_{\text{s}}}}\\
&=\left(\frac{M-S}{\sum_{f=1}^M \left(P_r(f)\right)^{\frac{-1}{n_{\text{s}}-1}}}\right)^{n_{\text{s}}}\left(\sum_{f=1}^M (P_r(f))^{\frac{-1}{n_{\text{s}}-1}}\right)=(M-S)^{n_{\text{s}}}\left(\sum_{f=1}^M (P_r(f))^{\frac{-1}{n_{\text{s}}-1}}\right)^{-(n_{\text{s}}-1)}\\
&=(M-S)^{n_{\text{s}}}\left(\sum_{f=1}^M \left(\frac{(f+q)^{-\gamma}}{H(1,M,\gamma,q)}\right)^{\frac{-1}{n_{\text{s}}-1}}\right)^{-(n_{\text{s}}-1)}=\frac{(M-S)^{n_{\text{s}}}}{H(1,M,\gamma,q)}\cdot\frac{1}{\left(\sum_{f=1}^M (f+q)^{\frac{\gamma}{n_{\text{s}}-1}}\right)^{n_{\text{s}}-1}}\\
&\stackrel{(a)}{\geq}\frac{(M-S)^{n_{\text{s}}}}{\frac{1}{1-\gamma}\left[(M+q)^{1-\gamma}-(1+q)^{1-\gamma}\right]+(1+q)^{-\gamma}}\cdot\frac{1}{\left(\int_1^{M+1}(x+q)^{\frac{\gamma}{\phi}}dx\right)^{\phi}}\\
&=\frac{(M-S)^{n_{\text{s}}}}{\frac{1}{1-\gamma}\left[(M+q)^{1-\gamma}-(1+q)^{1-\gamma}\right]+(1+q)^{-\gamma}}\cdot \frac{1}{\left(\frac{1}{\frac{\gamma}{\phi}+1}\right)^{\phi}\left[(M+q+1)^{\frac{\gamma}{\phi}+1}-(q+1)^{\frac{\gamma}{\phi}+1}\right]^{\phi}}\\
&=\frac{(1-\gamma)(M-S)^{n_{\text{s}}}}{(M+q)^{1-\gamma}-(1+q)^{1-\gamma}+(1-\gamma)(1+q)^{-\gamma}}\cdot\left(1-\frac{\gamma}{\phi+\gamma}\right)^{-\phi}\cdot\left[(M+q+1)^{\frac{\gamma}{\phi}+1}-(q+1)^{\frac{\gamma}{\phi}+1}\right]^{-\phi}\\
&=\frac{(1-\gamma)(M-S)^{n_{\text{s}}}}{(M+q)^{1-\gamma}-(1+q)^{1-\gamma}+(1-\gamma)(1+q)^{-\gamma}}\cdot e^{\gamma}\cdot\left[(M+q+1)^{\frac{\gamma}{\phi}+1}-(q+1)^{\frac{\gamma}{\phi}+1}\right]^{-\phi}\\
&=\frac{(1-\gamma)e^{\gamma}}{(M+q)^{1-\gamma}-(1+q)^{1-\gamma}}\cdot\left(\frac{M-S}{M}\right)^{n_{\text{s}}}\cdot (M)^{1-\gamma}\cdot\left[\left(\frac{M+q+1}{M}\right)^{\frac{\gamma}{\phi}+1}-\left(\frac{q+1}{M}\right)^{\frac{\gamma}{\phi}+1}\right]^{-\phi}+o(1),
\end{aligned}
\end{equation}
where $(a)$ is due to Lemma 1 of \cite{lee2019throughput}. Recall that $n_{\text{s}}=\rho' M$ and $D=\frac{q}{M}$. It follows from (\ref{eq:Lemma5_1}) that
\begin{equation}
\begin{aligned}\label{eq:Lemma5_2}
&p_{\text{miss}}(n_{\text{s}})\geq  \frac{(1-\gamma)e^{\gamma}\left(1-\frac{S}{M}\right)^{\rho' M}}{(1+D)^{1-\gamma}-(D+1/M)^{1-\gamma}}\left[\left(1+D+\frac{1}{M}\right)^{\frac{\gamma}{\phi}+1}-\left(D+\frac{1}{M}\right)^{\frac{\gamma}{\phi}+1}\right]^{-\phi}+o(1)\\
&=\frac{(1-\gamma)e^{-(S\rho'-\gamma)}}{(1+D)^{1-\gamma}-(D)^{1-\gamma}}\left[\left(1+D\right)^{\frac{\gamma}{\phi}+1}-\left(D\right)^{\frac{\gamma}{\phi}+1}\right]^{-\phi}+o(1)
\end{aligned}
\end{equation}
Finally, by using (\ref{eq:Lemma5_2}) and by letting $\rho'=\Theta(1)$, but $\rho'$ is arbitrarily large, we obtain $p_{\text{miss}}(n_{\text{s}})\geq \epsilon_1'(\rho')$, where $\epsilon_1'(\rho')$ can be arbitrarily small. Similarly, by letting  $\rho'=\omega(1)$, we obtain $p_{\text{miss}}(n_{\text{s}})\geq (1-\gamma)e^{-(S\rho'-\gamma)}=o(1)$. This completes the proof of Lemma 5.

\section{Proof of the Uniformly Random Matching}

When a user requests file $f$ and $\mathcal{V}_f$ is not empty, it needs to pick up a user uniformly random in $\mathcal{V}_f$ as the source. Then, since every user caches files according to the same randomized caching policy, the uniformly random selection of a user in $\mathcal{V}_f$ indicates the uniformly random selection of a user in the cluster, because the probabilities for the users to be included in $\mathcal{V}_f$ are the same. On the other hand, if $\mathcal{V}_f$ is empty, the user directly pick a user in the cluster uniformly at random as the source. Combining above statements with the fact that users request files according to the same distribution, we conclude that the proposed scheme in Sec. III.A matches users in the cluster uniformly at random.



\bibliographystyle{IEEEtran}
\bibliography{Thesis_2019_Bib__2020_03_23_}

\end{document}